\documentclass[11pt,titlepage]{article} 
\usepackage[sort&compress, numbers]{natbib}
\usepackage{amsfonts}
\usepackage{amsmath}
\usepackage{amssymb}
\usepackage{url}
\usepackage{tipa}
\usepackage{caption}
\DeclareCaptionLabelFormat{bf-parens}{\textbf{Supplementary #1 S#2}}
\usepackage{graphicx,rotating}
\usepackage{setspace}
\usepackage[table]{xcolor}
\usepackage{pdfpages}

\newcommand{\lachance}[0]{Lachance et al. [20] }
\newcommand{\smartpca}[0]{[22] }
\newcommand{\rolloff}[0]{[27] }
\newcommand{\treemix}[0]{[36] }
\newcommand{\schuster}[0]{[18] }
\newcommand{\henn}[0]{[17] }
\newcommand{\admixtools}[0]{[19] }
\newcommand{\denisova}[0]{[21] }
\newcommand{\tishkoff}[0]{[16] }
\newcommand{\wallace}[0]{[23] }
\newcommand{\kinahan}[0]{[33] }
\newcommand{\india}[0]{[26] }
\usepackage{etoolbox}

\makeatletter
\apptocmd{\thebibliography}{\global\c@NAT@ctr 38\relax}{}{}
\makeatother

\begin{document}

\includepdf[pages={1-16}]{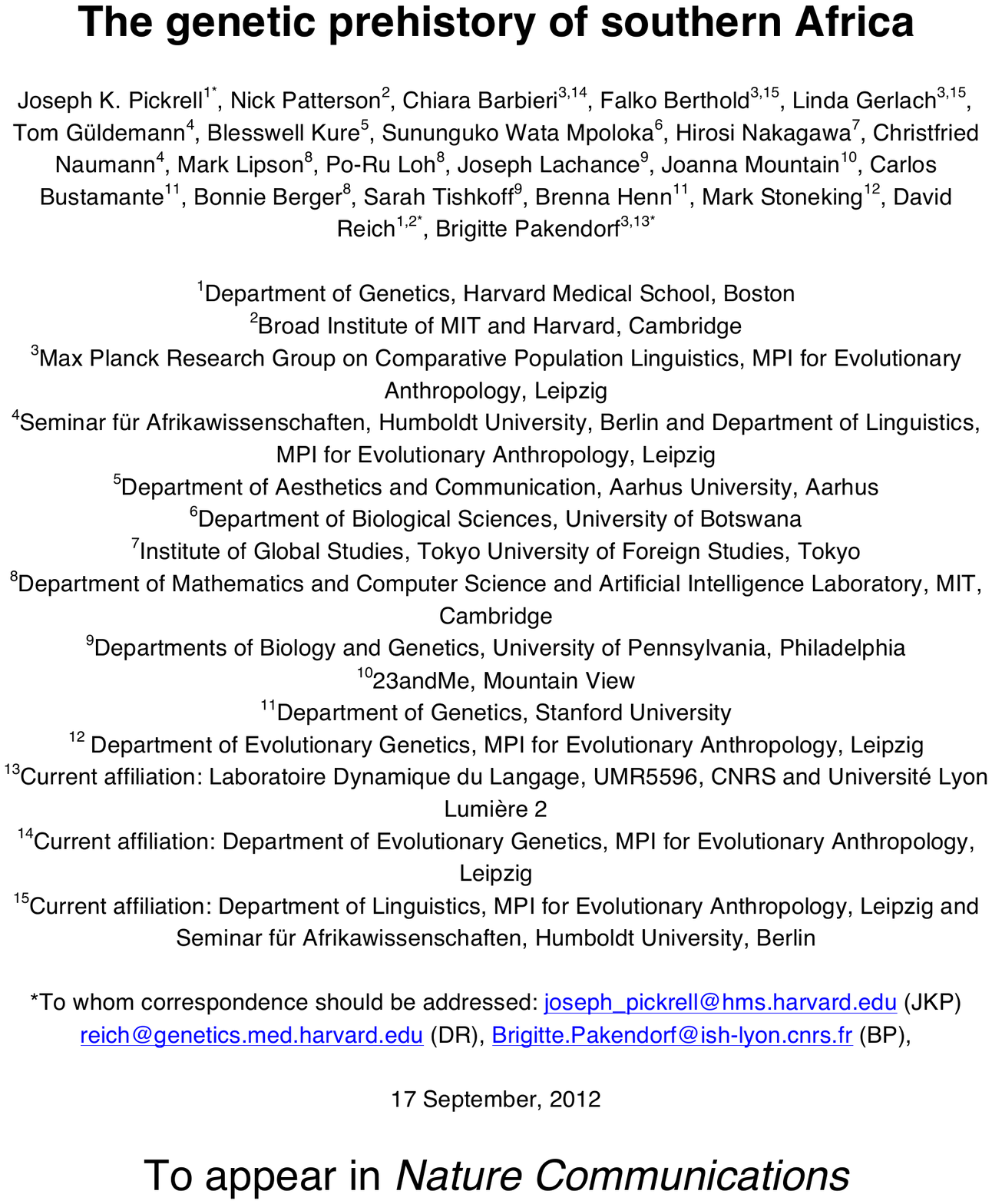}
\clearpage

\title{Supplementary Information for: The genetic prehistory of southern Africa}
 \small \author{Joseph K. Pickrell$^{1,\star}$, Nick Patterson$^2$, Chiara Barbieri$^{3,14}$, Falko Berthold$^{3,15}$,   \\
Linda Gerlach$^{3,15}$, Tom G\"{u}ldemann$^4$, Blesswell Kure$^{5}$, Sununguko Wata Mpoloka$^6$, \\
 Hirosi Nakagawa$^7$, Christfried Naumann$^4$, Mark Lipson$^8$, Po-Ru Loh$^8$, Joseph Lachance$^{9}$,\\
   Joanna L. Mountain$^{10}$, Carlos D. Bustamante$^{11}$, Bonnie Berger$^8$, Sarah Tishkoff$^{9}$, \\
  Brenna M. Henn$^{11}$, Mark Stoneking$^{12}$, David Reich$^{1, 2, \star}$, Brigitte Pakendorf$^{3,13, \star}$\\
  \\
  \small $^1$ Department of Genetics, Harvard Medical School, Boston\\
  \small $^2$ Broad Institute of MIT and Harvard, Cambridge\\
   \small $^3$ Max Planck Research Group on Comparative Population Linguistics, \\ 
  \small MPI for Evolutionary Anthropology, Leipzig \\
  \small $^4$ Seminar f\"ur Afrikawissenschaften, Humboldt University, Berlin and \\
 \small Department of Linguistics, MPI for Evolutionary Anthropology, Leipzig\\
   \small $^{5}$ Department of Aesthetics and Communication, Aarhus University, Aarhus\\
    \small $^6$ Department of Biological Sciences, University of Botswana \\
 \small $^7$ Institute of Global Studies, Tokyo University of Foreign Studies, Tokyo\\
  \small $^8$ Department of Mathematics and Computer Science and Artificial Intelligence Laboratory, MIT, Cambridge\\
  \small $^{9}$ Departments of Biology and Genetics, University of Pennsylvania, Philadelphia\\
 \small $^{10}$ 23andMe, Inc., Mountain View\\
 \small $^{11}$ Department of Genetics, Stanford University, Palo Alto\\
 \small $^{12}$ Department of Evolutionary Genetics, MPI for Evolutionary Anthropology, Leipzig\\
 \small $^{13}$ Current affiliation: Laboratoire Dynamique du Langage, UMR5596, CNRS and Universit\'{e} Lyon Lumi\`{e}re 2\\
  \small $^{14}$ Current affiliation: Department of Evolutionary Genetics, MPI for Evolutionary Anthropology, Leipzig\\
  \small $^{15}$ Current affiliation: Department of Linguistics, MPI for Evolutionary Anthropology, Leipzig\\
  \small and Seminar f\"ur Afrikawissenschaften, Humboldt University, Berlin \\
  \small $^\star$ To whom correspondence should be addressed:  Brigitte.Pakendorf@ish-lyon.cnrs.fr (BP)\\ 
\small reich@genetics.med.harvard.edu (DR), joseph\_pickrell@hms.harvard.edu (JP)
 }
 
\maketitle
\clearpage
\tableofcontents
\clearpage
\section{Supplementary Figures}

\begin{figure}[h!b]
\begin{center}
\includegraphics[scale = 0.65, angle=90]{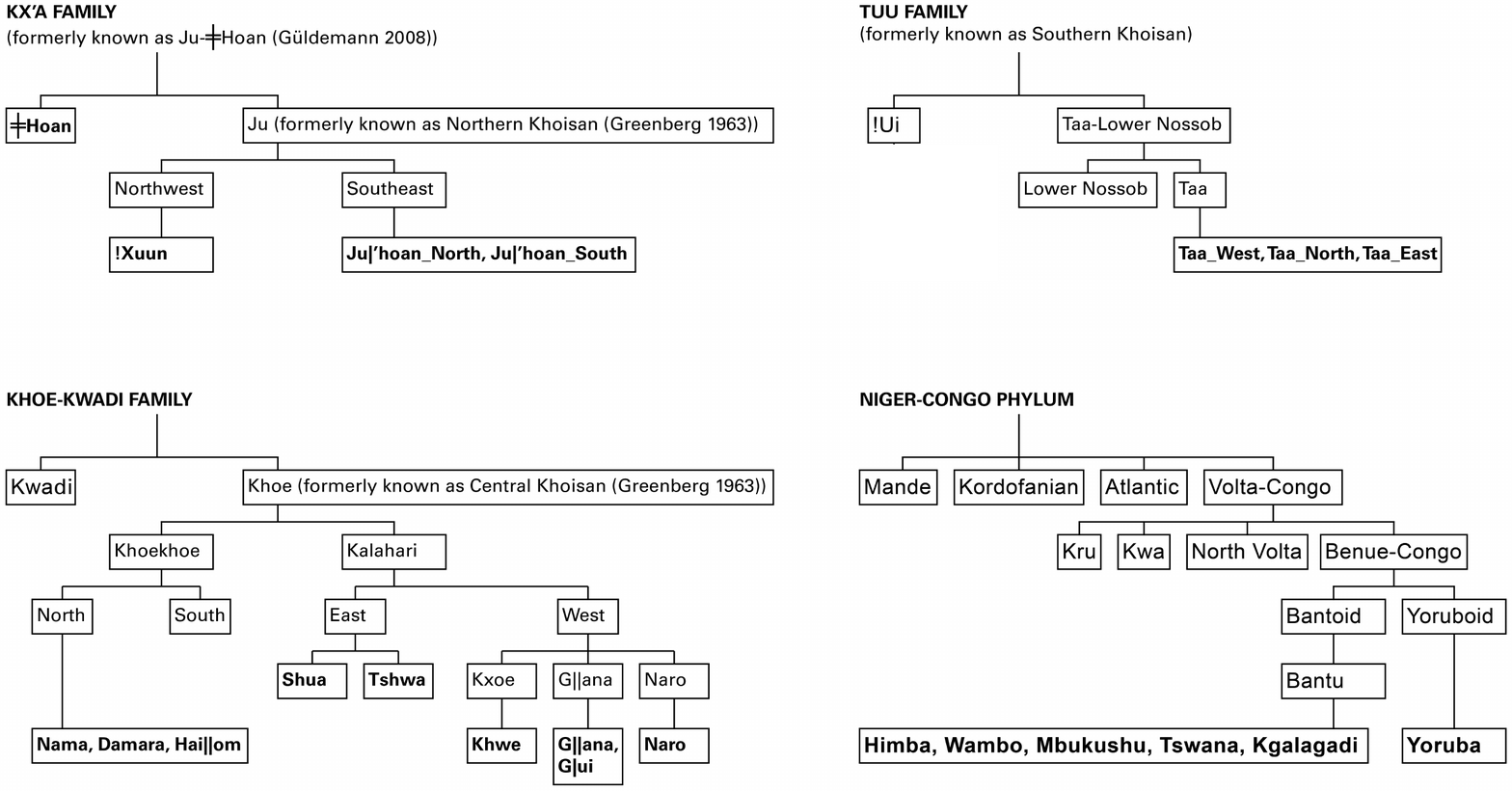}

\caption{\textbf{: Relationships between African languages spoken by populations in this study}. In bold are populations included in this study; the Hadza and Sandawe are not shown because they are linguistic isolates.}\label{fig_languages}

\end{center}
\end{figure}

\begin{figure}
\begin{center}
\includegraphics[scale = 1.4, angle=0]{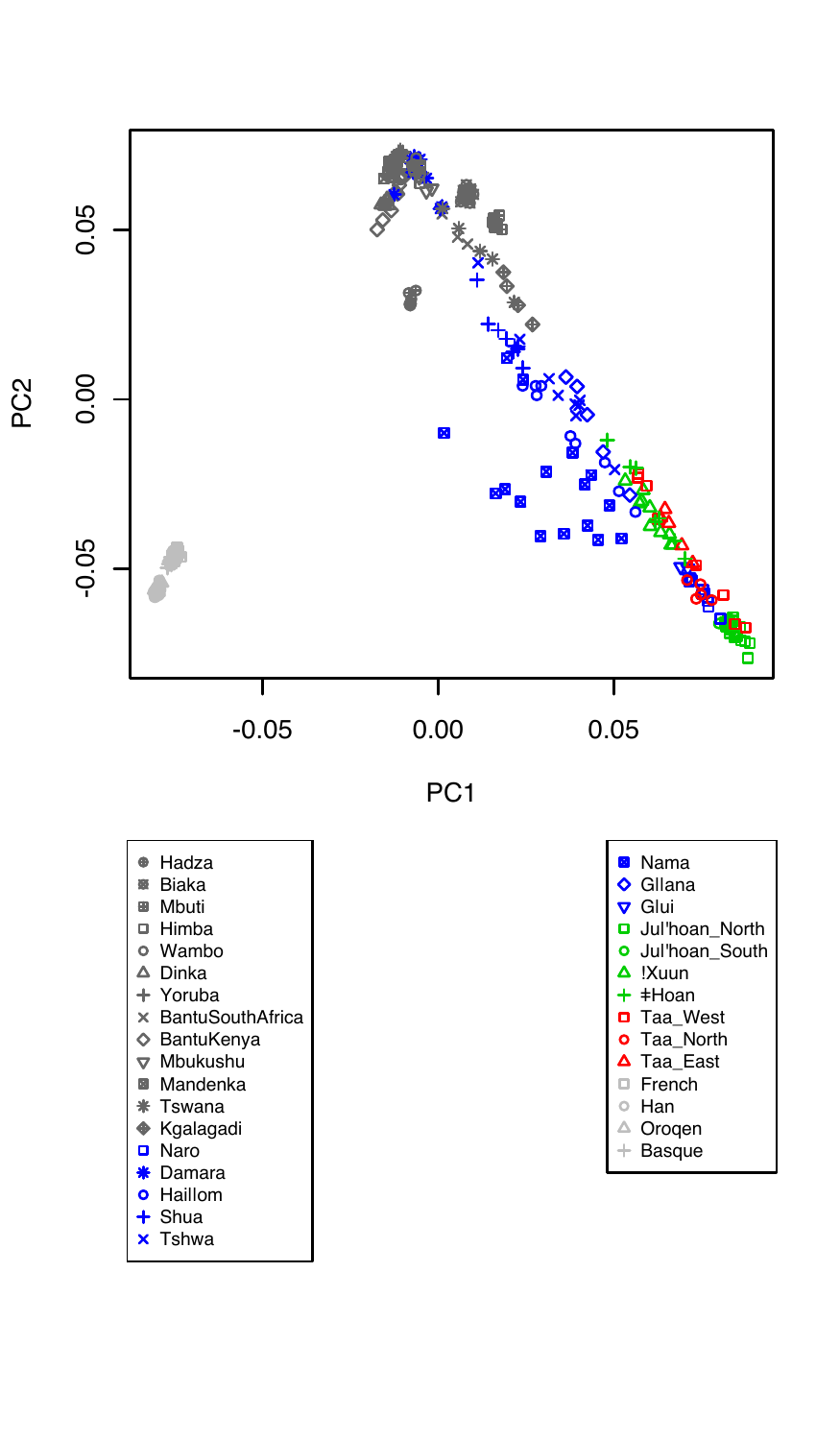}

\caption{\textbf{: PCA including non-African populations}. We performed principal component analysis on the genotype matrix of individuals using smartpca \smartpca using the SNPs ascertained in a Ju$|$'hoan\_North individual. Plotted are the positions of each individual along principal component axes one and two. The colors and symbols for each population are depicted in the legend.}\label{fig_pca_wnonaf}

\end{center}
\end{figure}
\clearpage

\begin{figure}
\begin{center}
\includegraphics[scale = 1.4]{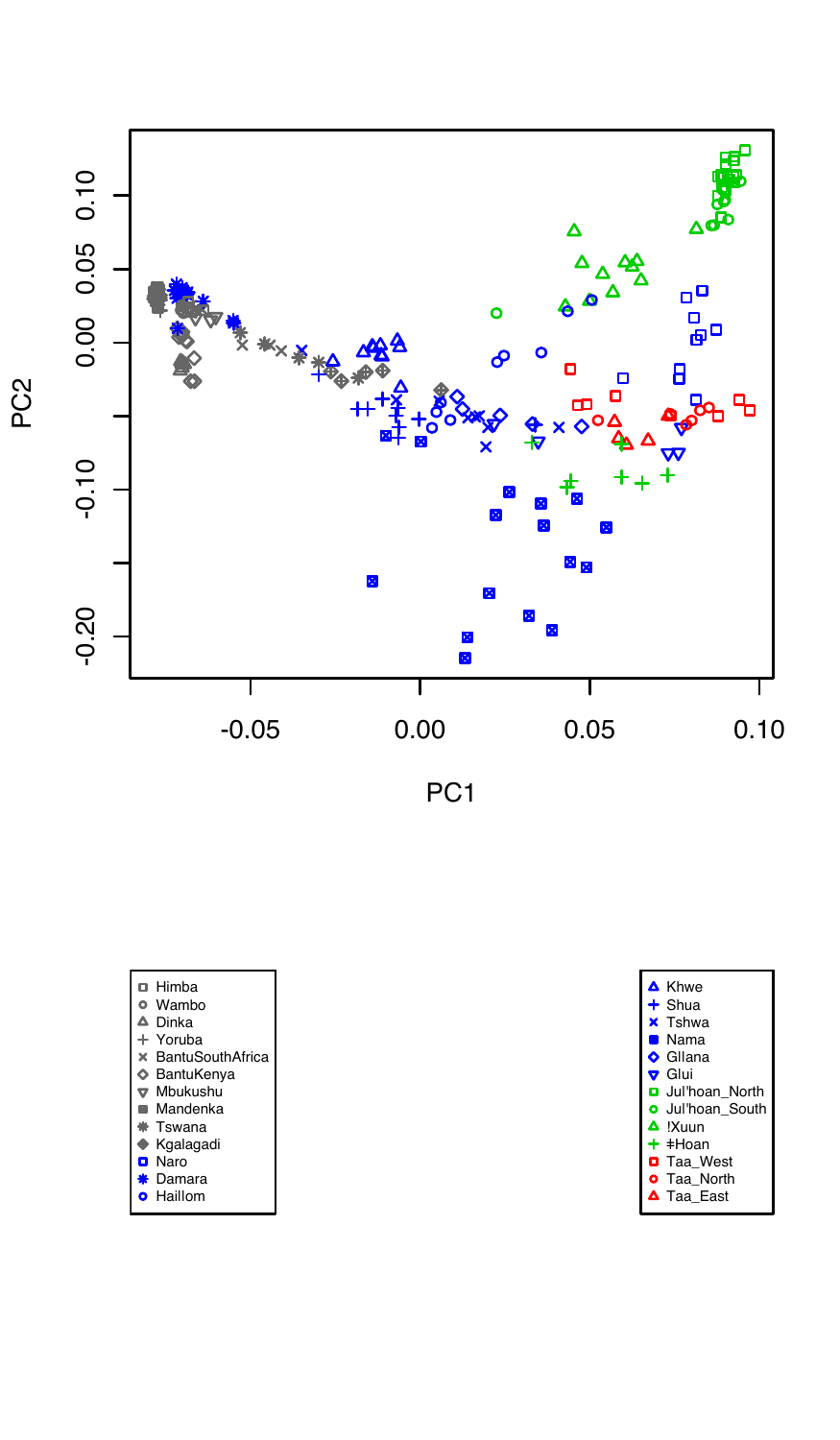}

\caption{\textbf{: PCA of African populations using all the SNPs on the chip.} Each individual is represented by a point, and the color and style of the point is displayed in the caption.}\label{fig_pca_allsnp}

\end{center}
\end{figure}

\begin{figure}
\begin{center}
\includegraphics[scale = 1]{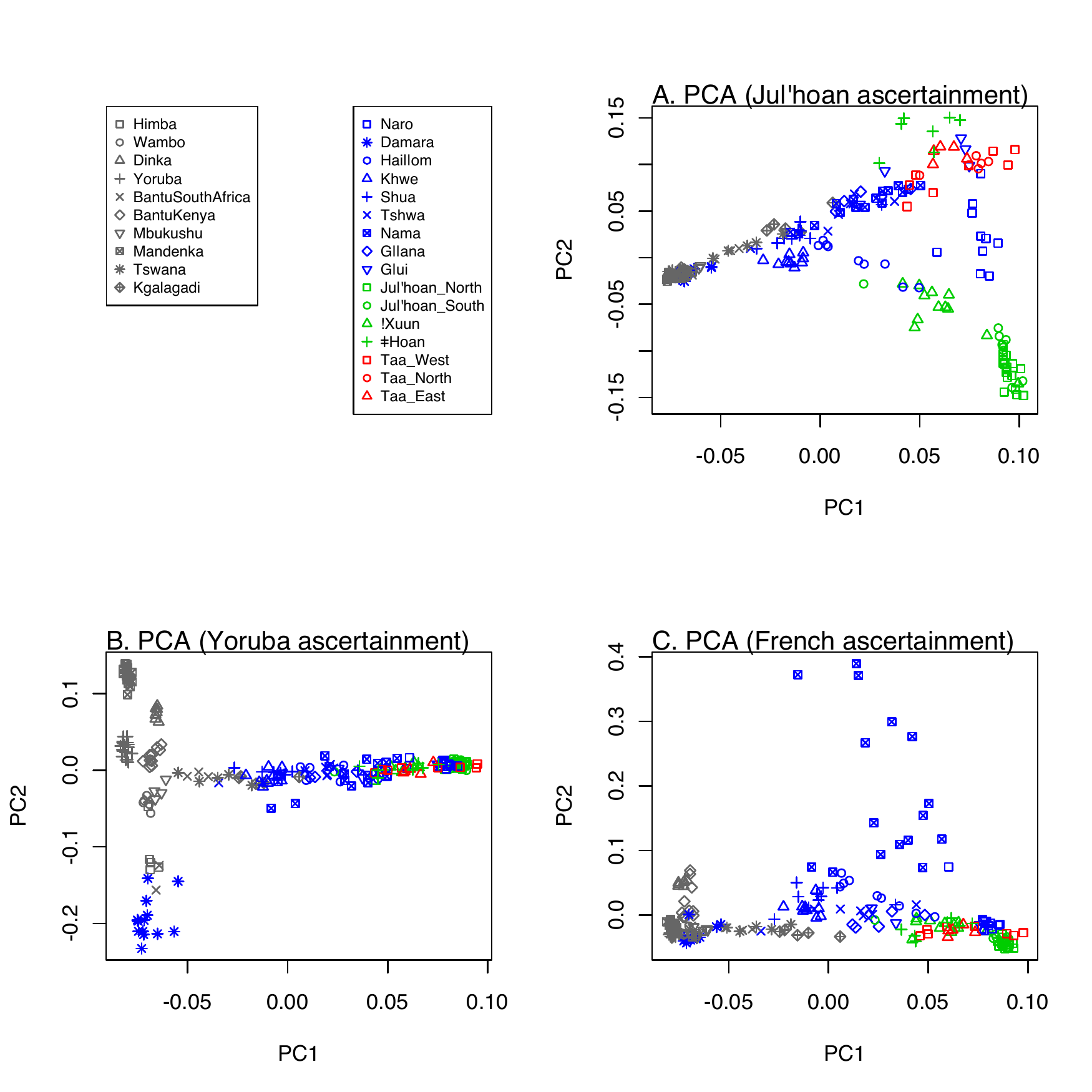}

\caption{\textbf{: PCA on SNPs from different ascertainment panels.} In each panel, each point represents an individual. The color and style of each point corresponds to the population of the individual as displayed in the legend. \textbf{A. Ju$|$'hoan\_North ascertainment.} This is same data presented in Figure 1B in the main text, but is included for comparison. \textbf{B. Yoruba ascertainment.} \textbf{C. French ascertainment}}\label{fig_pca_nooutliers}

\end{center}
\end{figure}

\begin{figure}
\begin{center}
\includegraphics[scale = 1.3]{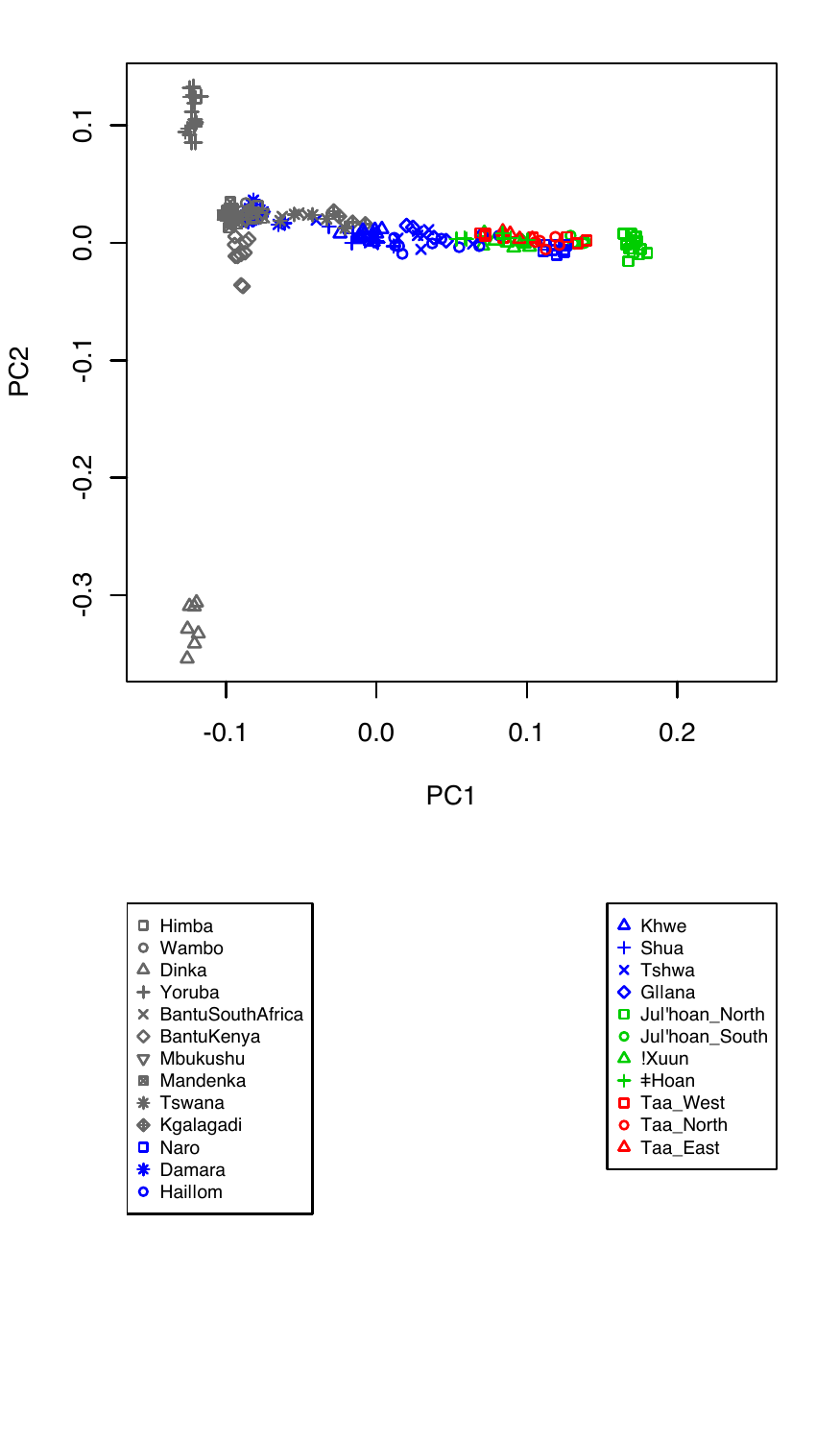}

\caption{\textbf{: PCA projection using Ju$|$'hoan\_North, Yoruba, and Dinka.} We identified principal components using only the Ju$|$'hoan\_North, Yoruba, and Dinka, then projected the other samples (excluding outliers) onto these axes. All Khoisan populations fall on a cline between the Ju$|$'hoan\_North and the neighboring Bantu-speaking populations. This is consistent with the variation in non-Khoisan admixture in these populations being due to variation in admixture with neighboring agriculturist populations.}\label{fig_pca_project}

\end{center}
\end{figure}

\begin{figure}
\begin{center}
\includegraphics[scale = 0.8]{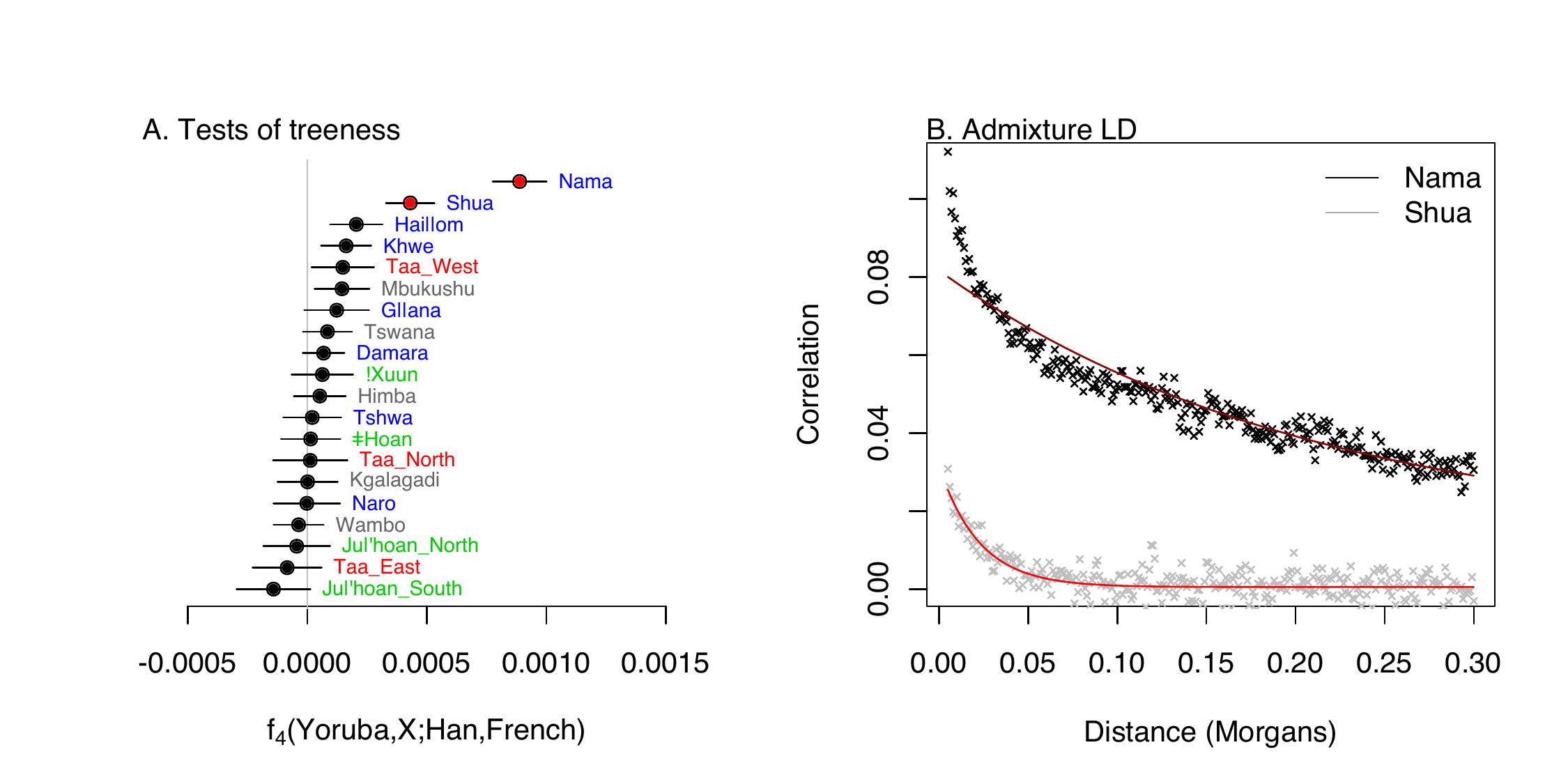}

\caption{\textbf{: The Nama have recent non-African ancestry. A. Four-population tests.} We performed four-population tests on the tree topology [[Yoruba,X],[Han,French]], where X represents any southern African population. Plotted is the value of the $f_4$ statistic when each southern African population is used. Error bars show a single standard error, and points in red have a Z-score greater than 3. \textbf{B. Admixture LD.} We ran ROLLOFF on the Nama and the Shua using the Ju$|$'hoan\_North and the French as the mixing populations. There is a clear decay in the Nama (the shift away from the x-axis is indicative of variable ancestry across individuals, which is visually apparent in Supplementary Figure S\ref{fig_pca_wnonaf}) and a less obvious decay in the Shua. The red lines show the fitted exponential curves.}\label{fig_mix_eu}

\end{center}
\end{figure}




\begin{figure}
\begin{center}
\includegraphics[scale = 1]{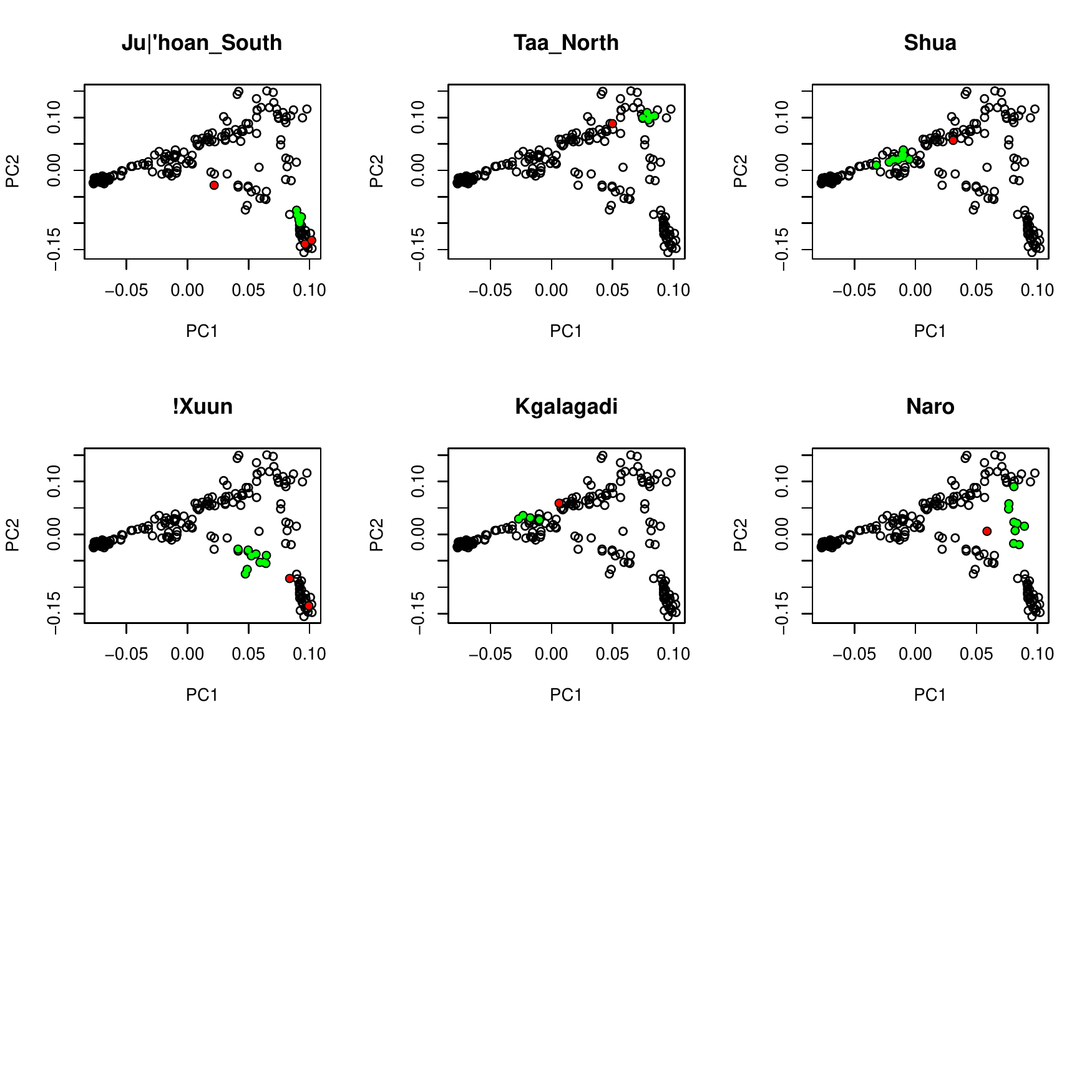}

\caption{\textbf{: Individuals excluded from populations.} Shown are the PCA plot in Figure 1 in the main text, with different populations highlighted. In red are the individuals we excluded, and in green those that were kept. See Supplementary Table S\ref{samp_table} for total sample sizes in each population. }\label{outliers}

\end{center}
\end{figure}

\begin{figure}
\begin{center}
\includegraphics[scale = 0.8,angle = 270]{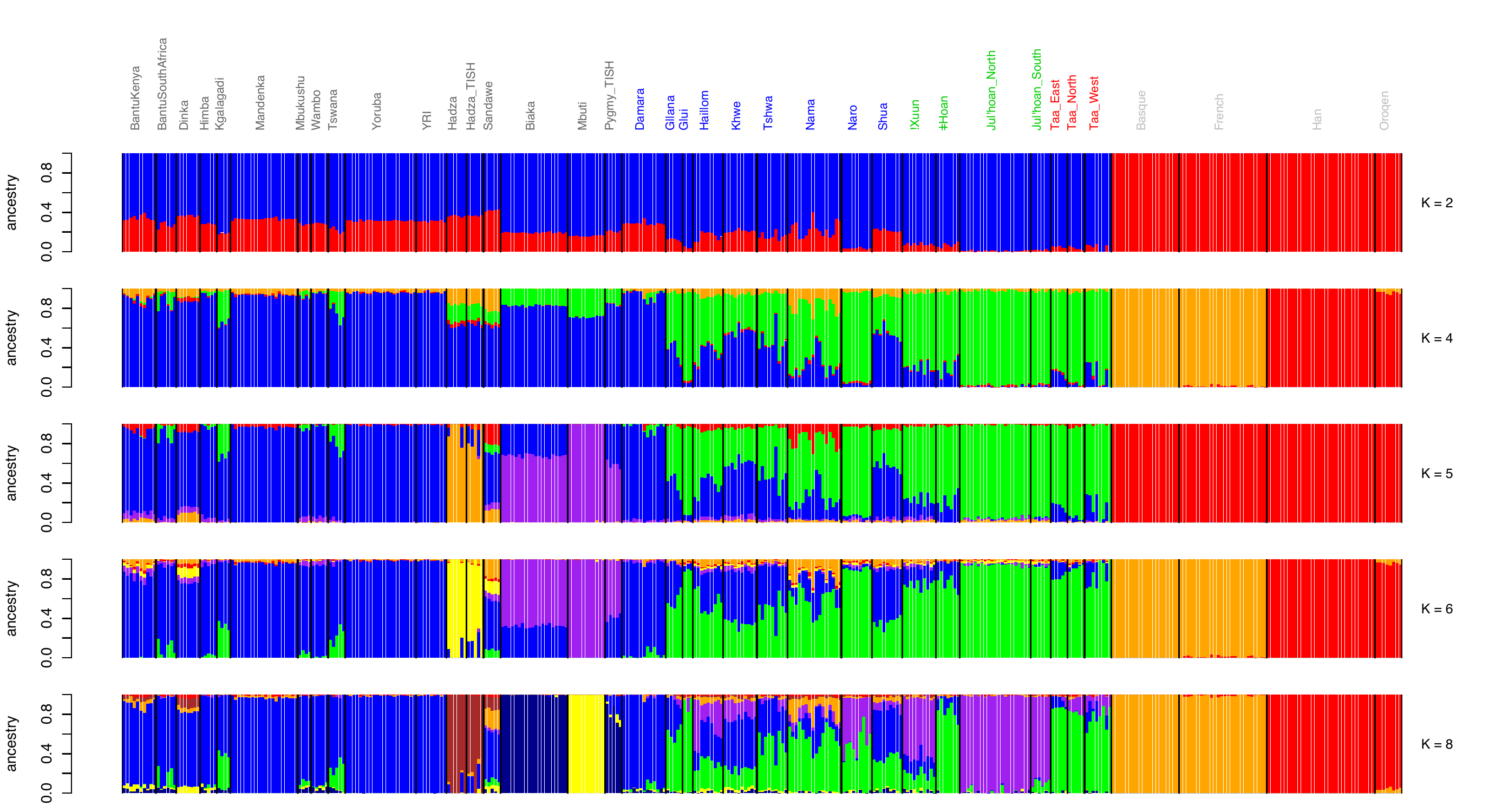}

\caption{\textbf{: Clustering analyses of the merged sequenced and genotyped samples.} We ran ADMIXTURE on all African individuals using different settings of $K$; shown are the resulting clusters. The populations merged from \lachance are YRI (compare to Yoruba), Hadza\_TISH (compare to Hadza), Pygmy\_TISH, and Sandawe.}\label{southaf_admix_wtish}

\end{center}
\end{figure}

\begin{figure}
\begin{center}
\includegraphics[scale = 1]{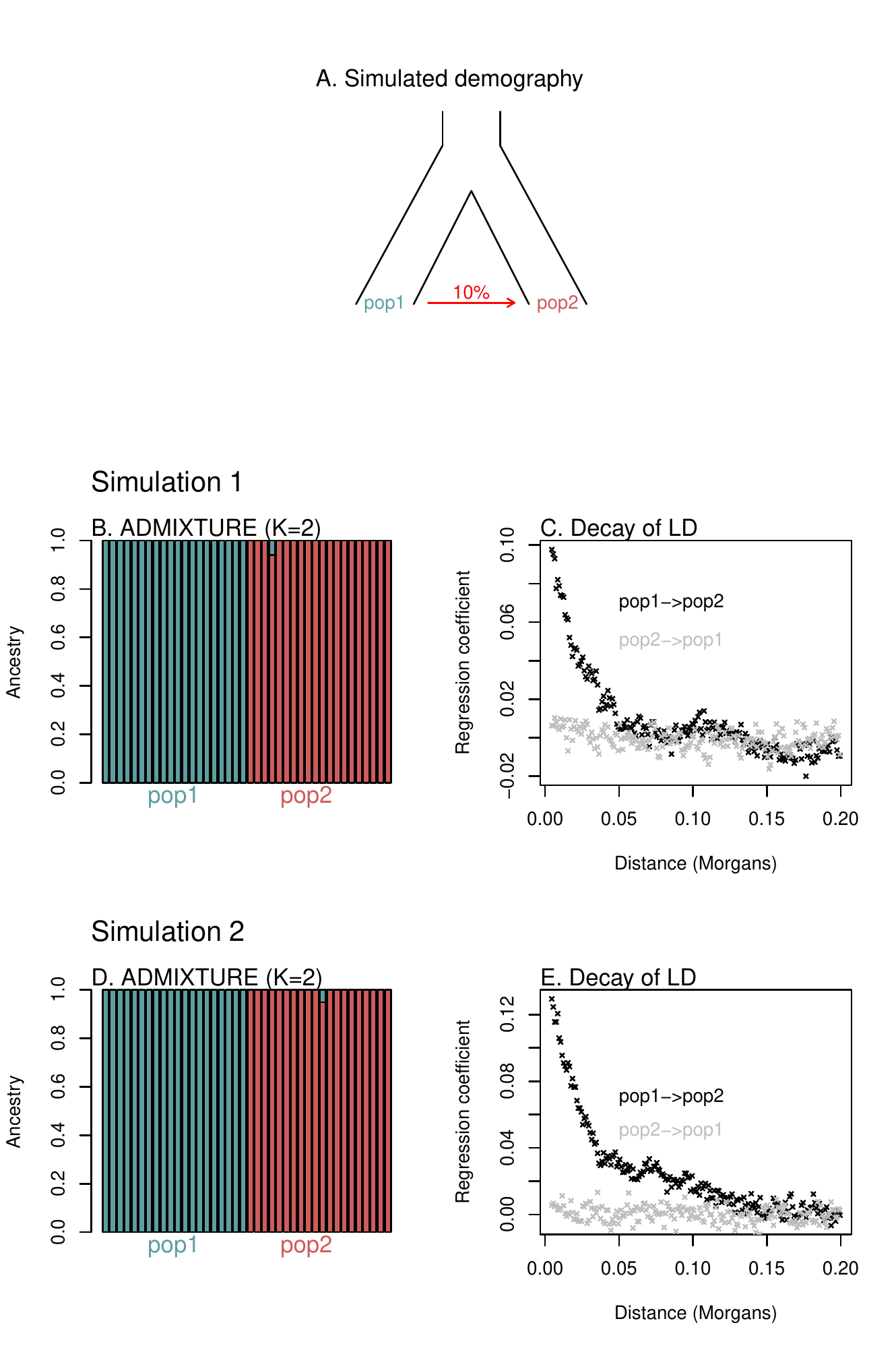}

\caption{\textbf{: LD information identifies previously undetectable admixture events}. We performed simulations of two populations, one of which admixed with the other 40 generations in the past (see Section \ref{ld_section} for details). Shown are results from two simulations. \textbf{A.} The simulated demography. \textbf{B,D.} Results from running ADMIXTURE on the simulated data. \textbf{C,E.} Results from the measure of LD decay described in Section \ref{ld_section}.}\label{admix_sim}

\end{center}
\end{figure}

\begin{figure}
\begin{center}
\includegraphics[scale = 0.8]{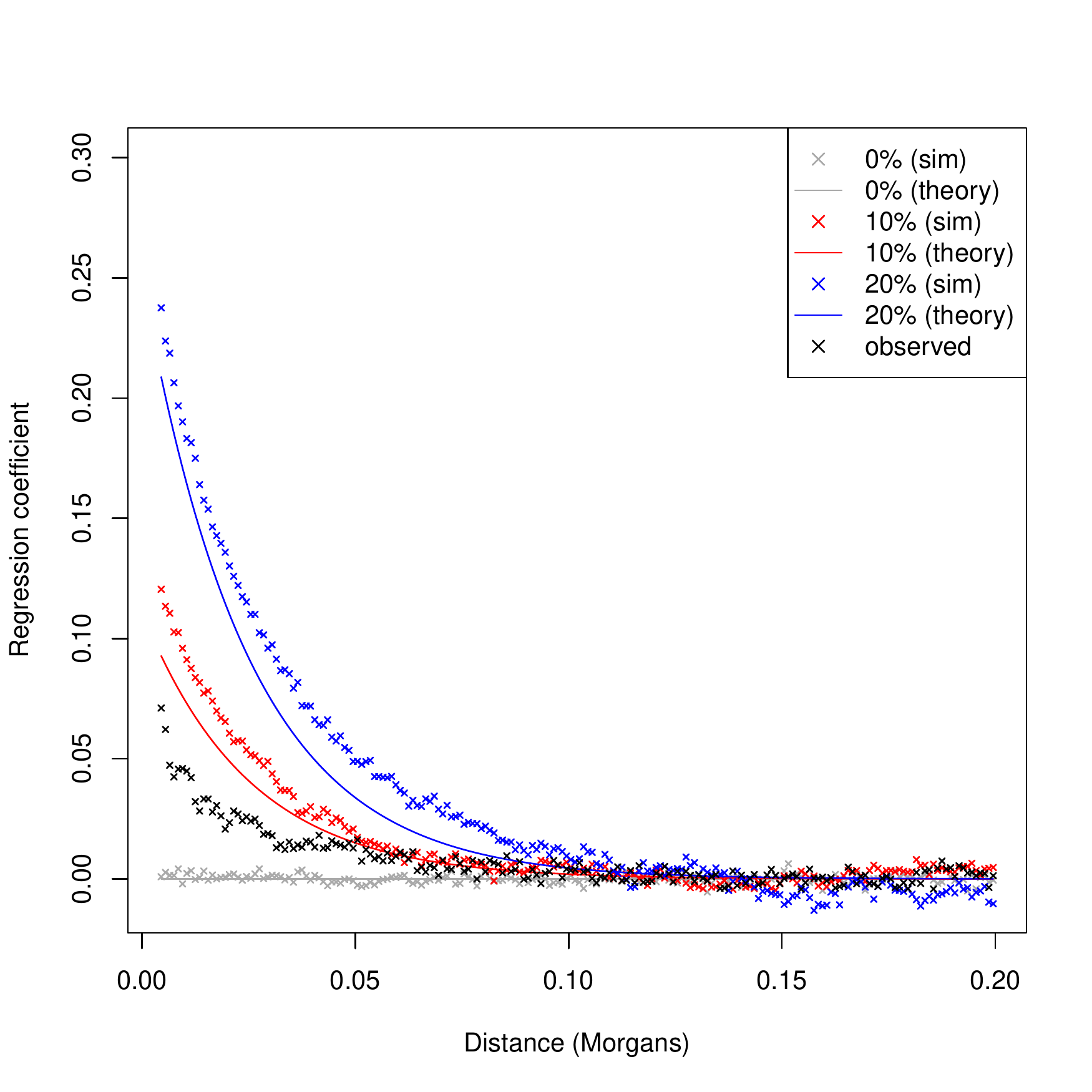}

\caption{\textbf{: Estimating mixture proportions from LD.} We simulated genetic data under different demographies including admixture (Supplementary Information), and then estimated admixture proportions using the method described in the text. The colored and grey points represent the decay curve obtained in simulations (each curve is the average of five simulations of 100 Mb), and the lines are the theoretical curves. In black is the data from the Ju$|$'hoan\_North and Yoruba, treating the Ju$|$'hoan\_North as admixed.}\label{rolloff_sim}

\end{center}
\end{figure}

\begin{figure}
\begin{center}
\includegraphics[scale = 1]{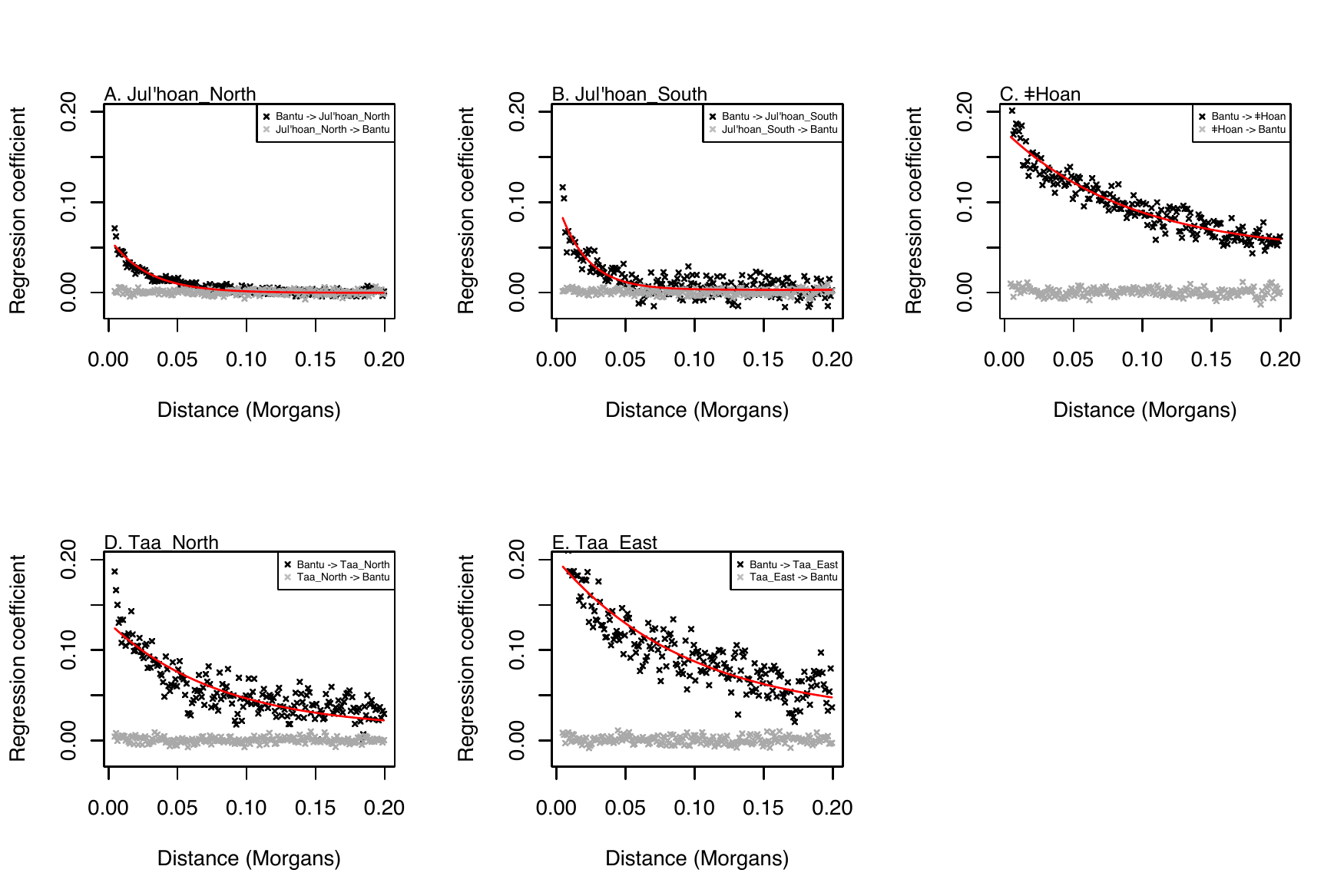}

\caption{\textbf{: Admixture LD in populations that pass the three-population test.} We measured the decay of admixture LD on the five Khoisan populations that show no evidence of admixture in three-population tests. The method is described in Section \ref{ld_section}. Each panel shows an individual population; panel \textbf{A}. is a version of Figure 2A from the main text with the y-axis modified to be the same as the other panels. In all cases, the non-Khoisan population used in the analysis is the Yoruba. In red is the fitted exponential curve. }\label{rolloff_noadmix}

\end{center}
\end{figure}

\begin{figure}
\begin{center}
\includegraphics[scale = 1]{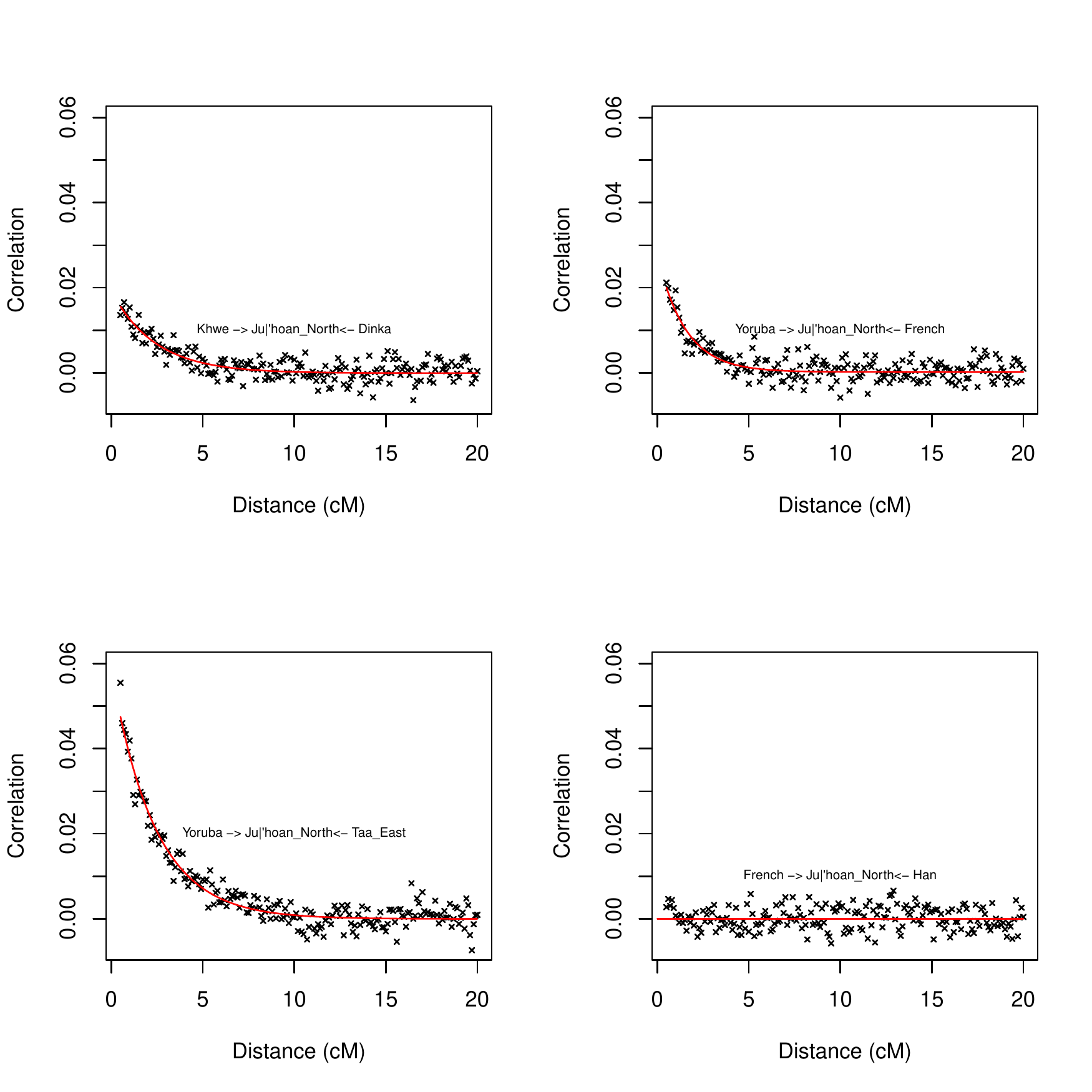}

\caption{\textbf{: ROLLOFF analysis of the Ju$|$'hoan\_North.} We explored the correlation between the decay of LD in the Ju$|$'hoan\_North and the divergence between other pairs of populations using ROLLOFF. At each pair of SNPs, we estimate the amount of LD in the Ju$|$'hoan\_North (as measured by a correlation in genotypes \rolloff) and the product of the differences in allele frequency between two reference populations. The reference populations in each panel are listed to either side of the Ju$|$'hoan\_North. We then calculate the correlation between these two values, binning pairs of SNPs by the genetic distance between them.  Each point is the value of this correlation (the y-axis) plotted against the genetic distance bin (the x-axis). A detectable curve suggests that the target population (in this case the Ju$|$'hoan) is admixed. Note a curve can be present even if the reference populations are quite distant from the true mixing populations. In this case, a curve is seen except when using two non-African populations as references. In red is the fitted exponential curve.}\label{rolloff_ju}

\end{center}
\end{figure}

\begin{figure}
\begin{center}
\includegraphics[scale = 1]{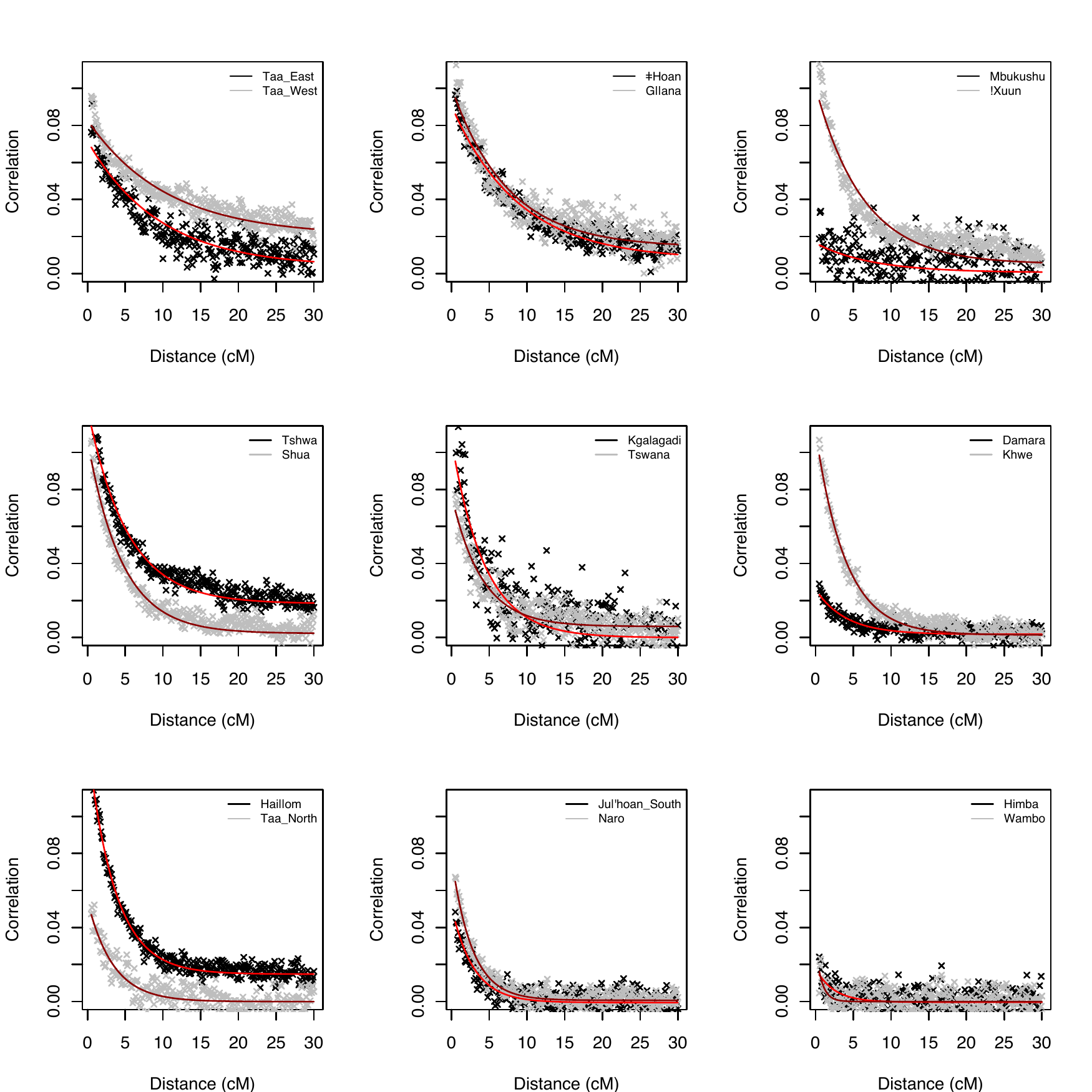}

\caption{\textbf{: ROLLOFF analysis of all southern African populations}. For each southern African population, we ran ROLLOFF \rolloff using the Ju$|$'hoan\_North and Yoruba as the mixing populations. The method is as described in Supplementary Figure S\ref{rolloff_ju} and the Supplementary Material. Shown are the resulting curves for each population; in red are the fitted exponential curves.}\label{rolloff}

\end{center}
\end{figure}

\clearpage

\begin{figure}
\begin{center}
\includegraphics[scale = 1]{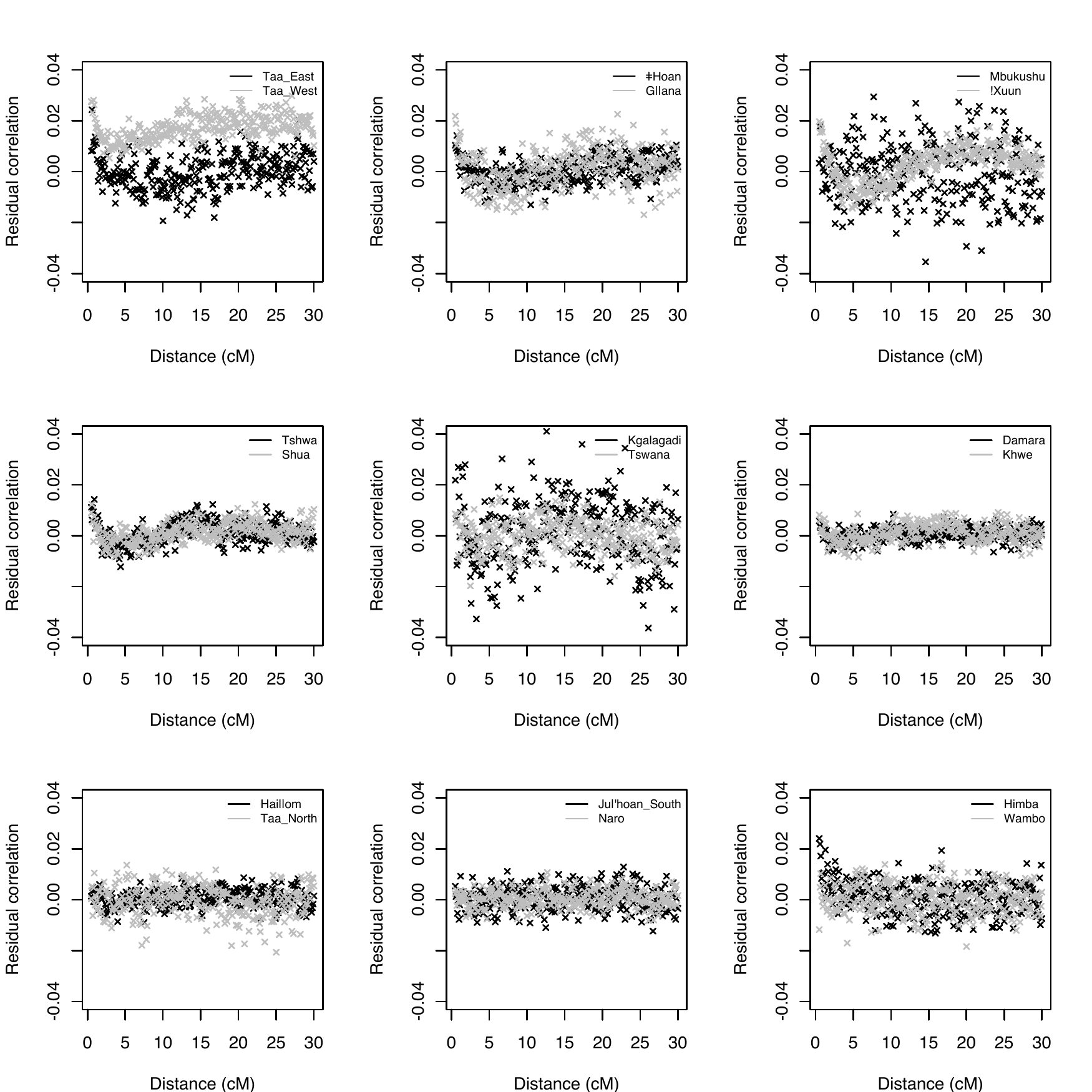}

\caption{\textbf{: Residuals from ROLLOFF analysis of all southern African populations}. Plotted are the residuals from the fit of the exponential curves for each population from Supplementary Figure S\ref{rolloff}. Residual correlation may indicate multiple waves of mixture in some populations.}\label{rolloff_resid}

\end{center}
\end{figure}

\begin{figure}
\begin{center}
\includegraphics[scale = 1]{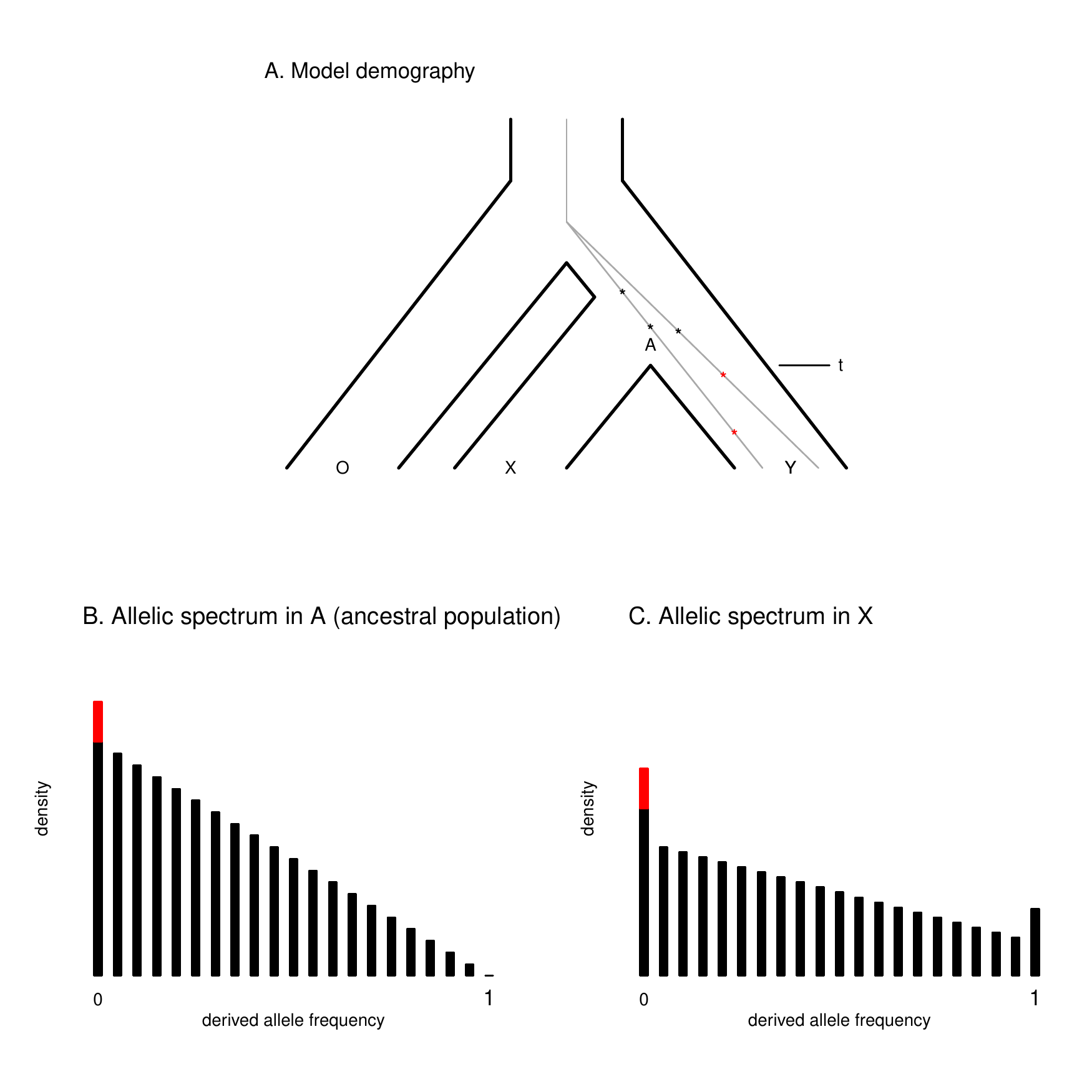}

\caption{\textbf{: Scheme for dating population splits. A. Demographic model}. Plotted is the demographic model used in our method for dating population split times. Populations are labeled in black, and the split time is denoted t. In grey is the history of the two chromosomes used for SNP ascertainment. Stars represent mutations, and are colored according to whether they arose before (black) or after (red) the population split. \textbf{B.} A hypothetical allelic spectrum in population $A$. The red peak at zero corresponds to the mutations that happened on the lineage to $Y$. \textbf{C.} The hypothetical allelic spectrum in $X$. Though alleles change frequency from $A$ to $X$, the size of the red component of the peak at zero stays constant. }\label{fig_split}
\end{center}
\end{figure}

\begin{figure}
\begin{center}
\includegraphics[scale = 0.6]{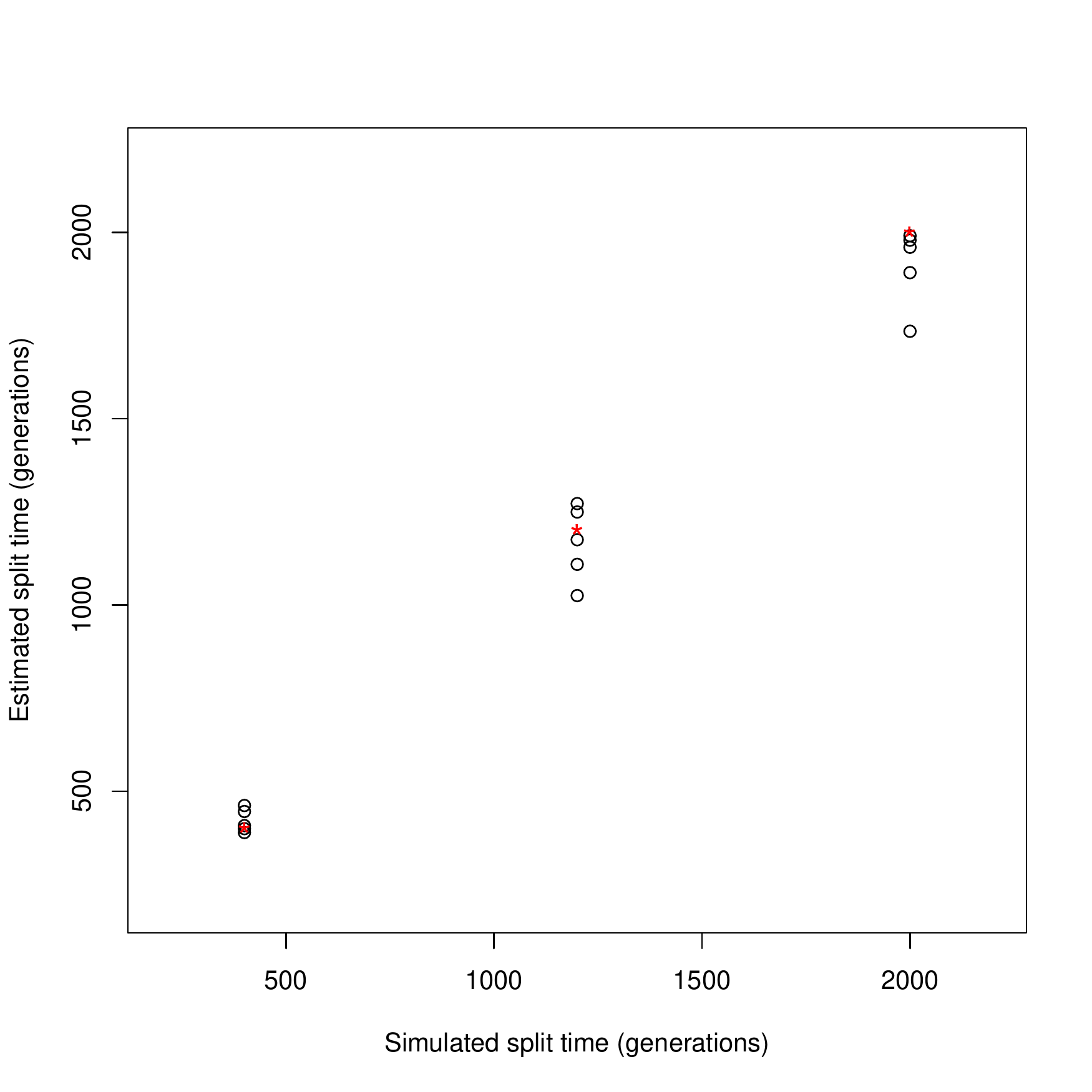}

\caption{\textbf{: Estimating split times in simulations without migration}. Shown are the simulated and inferred split times in simulations without migration. The red stars show the true simulated values, and the black points the estimates.}\label{fig_sim}
\end{center}
\end{figure}

\begin{figure}
\begin{center}
\includegraphics[scale = 1]{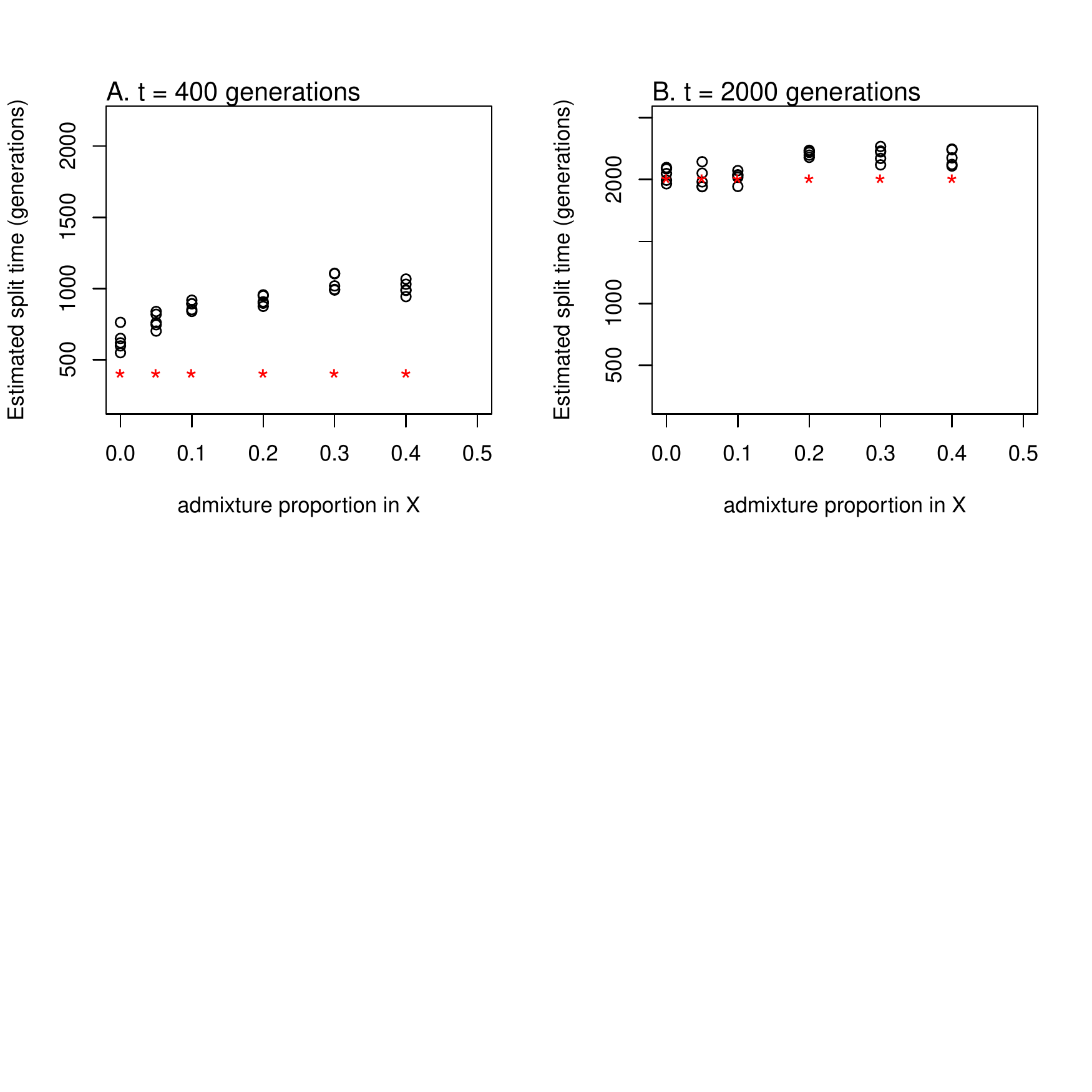}

\caption{\textbf{: Estimating split times in simulations with migration}. Shown are estimated split times between $X$ and $Y$ when both have experienced some level of admixture with an outgroup. The black points show estimated split times in the presence of admixture. The red stars show the true simulated values. In all simulations, population $Y$ has 5\% admixture from the outgroup that occurred 40 generations in the past, while population $X$ has variable levels of admixture (plotted on the x-axis).}\label{fig_sim_wmig}
\end{center}
\end{figure}

\begin{figure}
\begin{center}
\includegraphics[scale = 1]{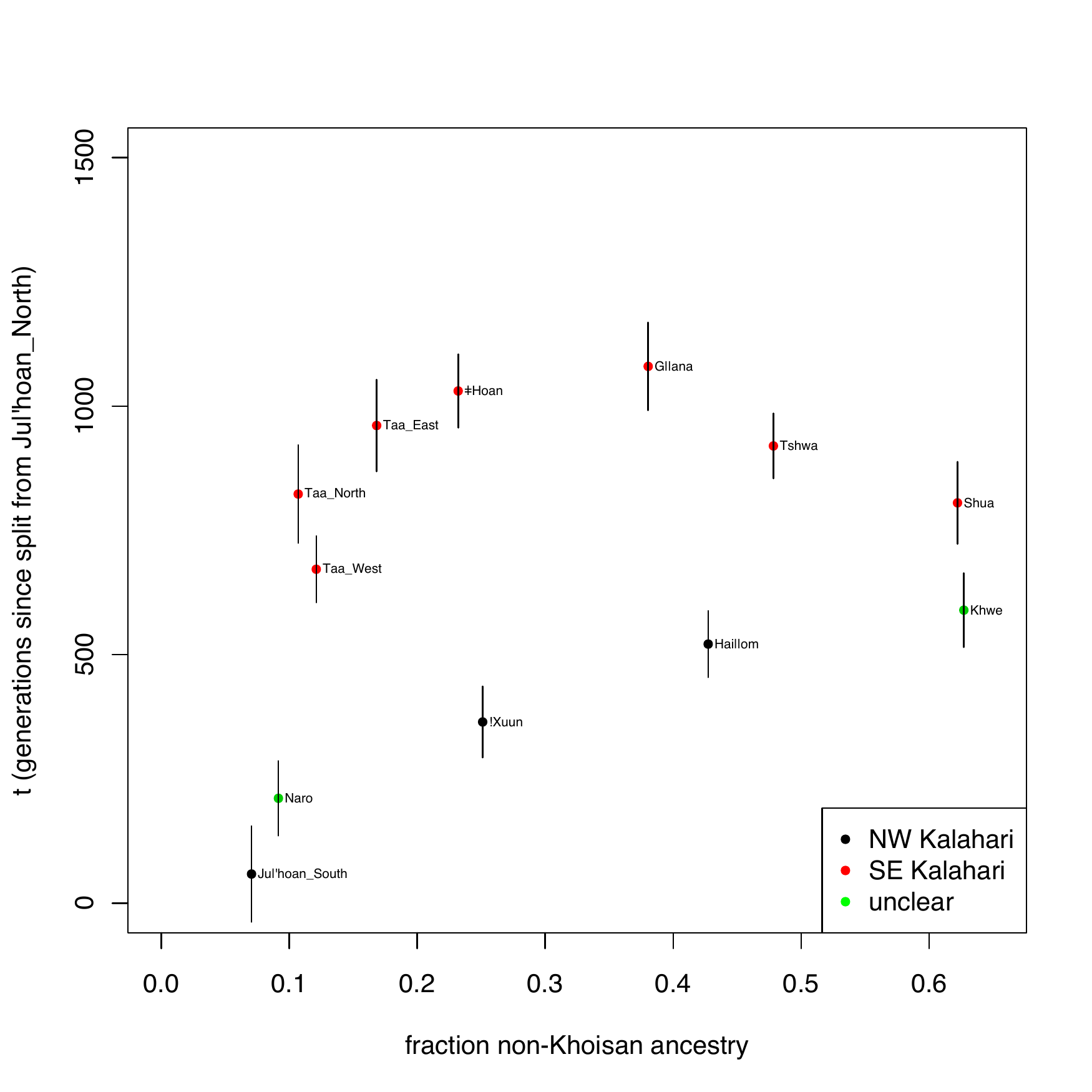}

\caption{\textbf{: Dating the split time of the Khoisan populations.} We plot the estimated times since each Khoisan population split from the Ju$|$'hoan\_North, as a function of their level of non-Khoisan-admixture. The populations included are all the southern African groups in Figure 3 in the main text. The errors bars are one standard error (not including the error in the estimate of $\tau$). Khoisan populations are colored according to whether they have strong evidence (from Figure 3 in the main text) as coming from the northwestern Kalahari cluster or the southeastern Kalahari cluster. Populations that have no clear grouping are colored in green. All split times are likely overestimated due to non-Khoisan admixture (see Supplementary Figure S\ref{fig_sim_wmig})}\label{fig_khoisan}
\end{center}
\end{figure}

\begin{figure}
\begin{center}
\includegraphics[scale = 1]{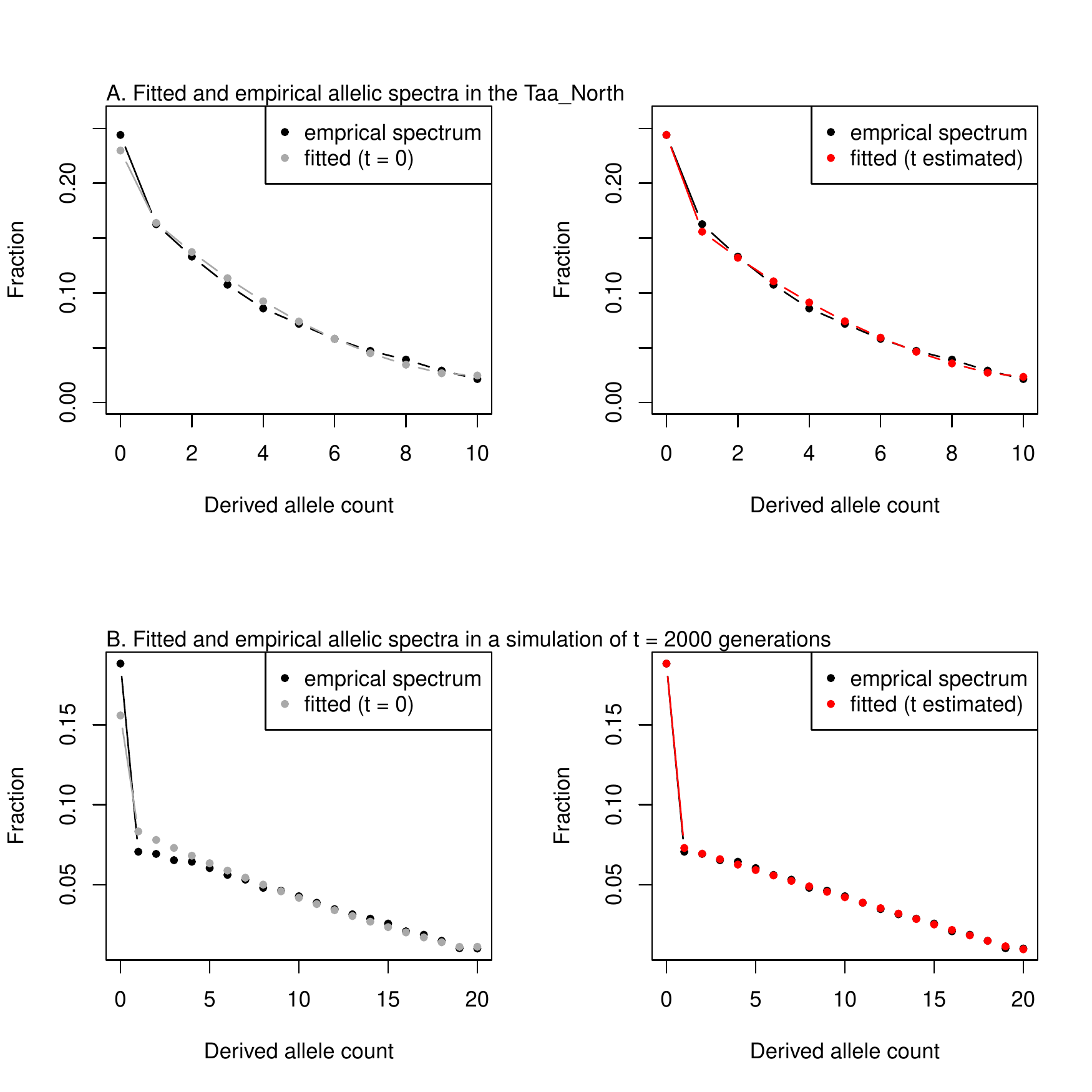}

\caption{\textbf{: Dating the split time of Taa\_North. A.} We plot the empirical allele frequency spectrum in the Taa\_North at SNPs ascertained in a single Ju$|$'hoan\_North individual (in black). For comparison we plot the fitted allelic spectra if we assume the split time between Taa\_North and the Ju$|$'hoan\_North is zero (in grey in the left panel) or if we allow the model to estimate the split time (in red in the right panel). Note that the empirical spectrum is non-linear, implying that the ancestral population was not of constant size. \textbf{B.} We plot the analogous spectra for a single simulation of a split time of 2,000 generations with no migration.}\label{fig_ntaa}
\end{center}
\end{figure}


\begin{figure}
\begin{center}
\includegraphics[scale = 1]{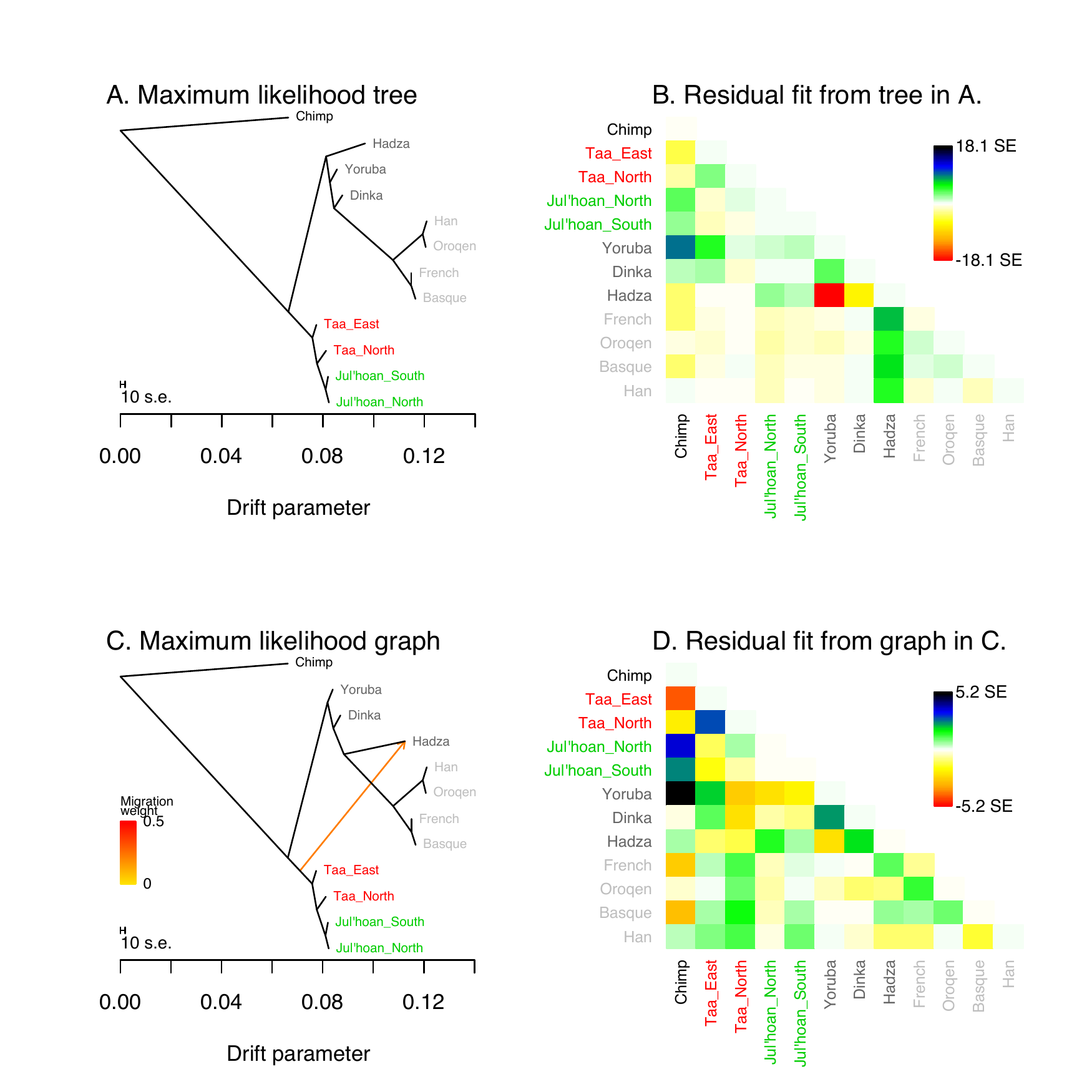}

\caption{\textbf{: \emph{TreeMix} analysis of the Hadza}. Shown is the maximum likelihood tree of populations including the Hadza (A.), the residual fit from this tree (B.), the inferred graph allowing for a single migration edge (C.), and the residual fit from this graph (D.). See Supplementary text for discussion. Note that the choice of which edge to the Hadza is called the ``migration" edge is arbitrary \treemix; for Figure 3 in the main text we force the non-Khoisan ancestry in the Hadza to be the ``migration" edge.}\label{treemix_hadza}

\end{center}
\end{figure}

\begin{figure}
\begin{center}
\includegraphics[scale = 1]{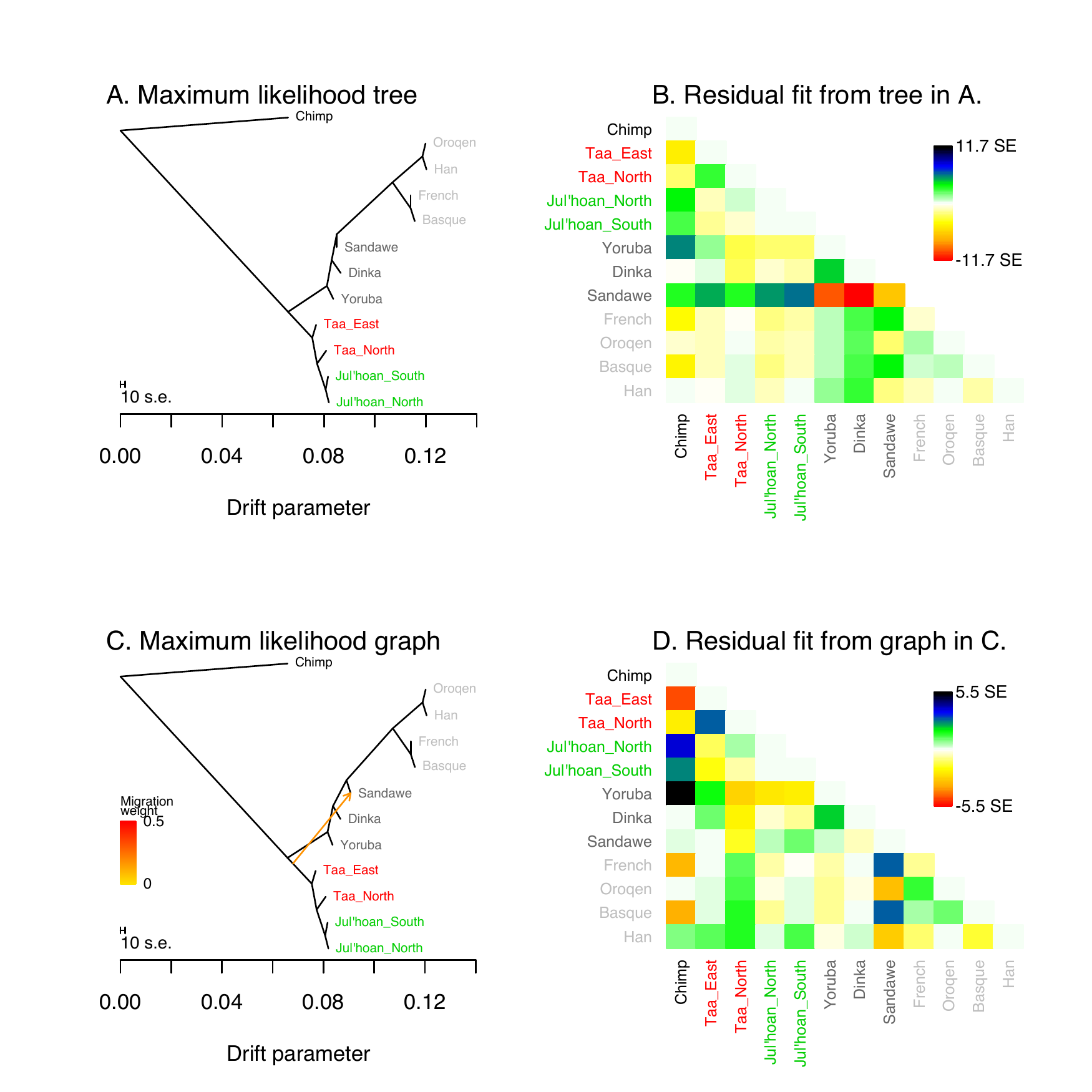}

\caption{\textbf{: \emph{TreeMix} analysis of the Sandawe.} Shown is the maximum likelihood tree of populations including the Sandawe individuals. (A.), the residual fit from this tree (B.), the inferred graph allowing for a single migration edge (C.), and the residual fit from this graph (D.). See Supplementary text for discussion. Note that the choice of which edge to the Sandawe is called the ``migration" edge is arbitrary \treemix; for Figure 3 in the main text we force the non-Khoisan ancestry in the Sandawe to be the ``migration" edge. }\label{treemix_sandawe}

\end{center}
\end{figure}

\begin{figure}
\begin{center}
\includegraphics[scale = 1]{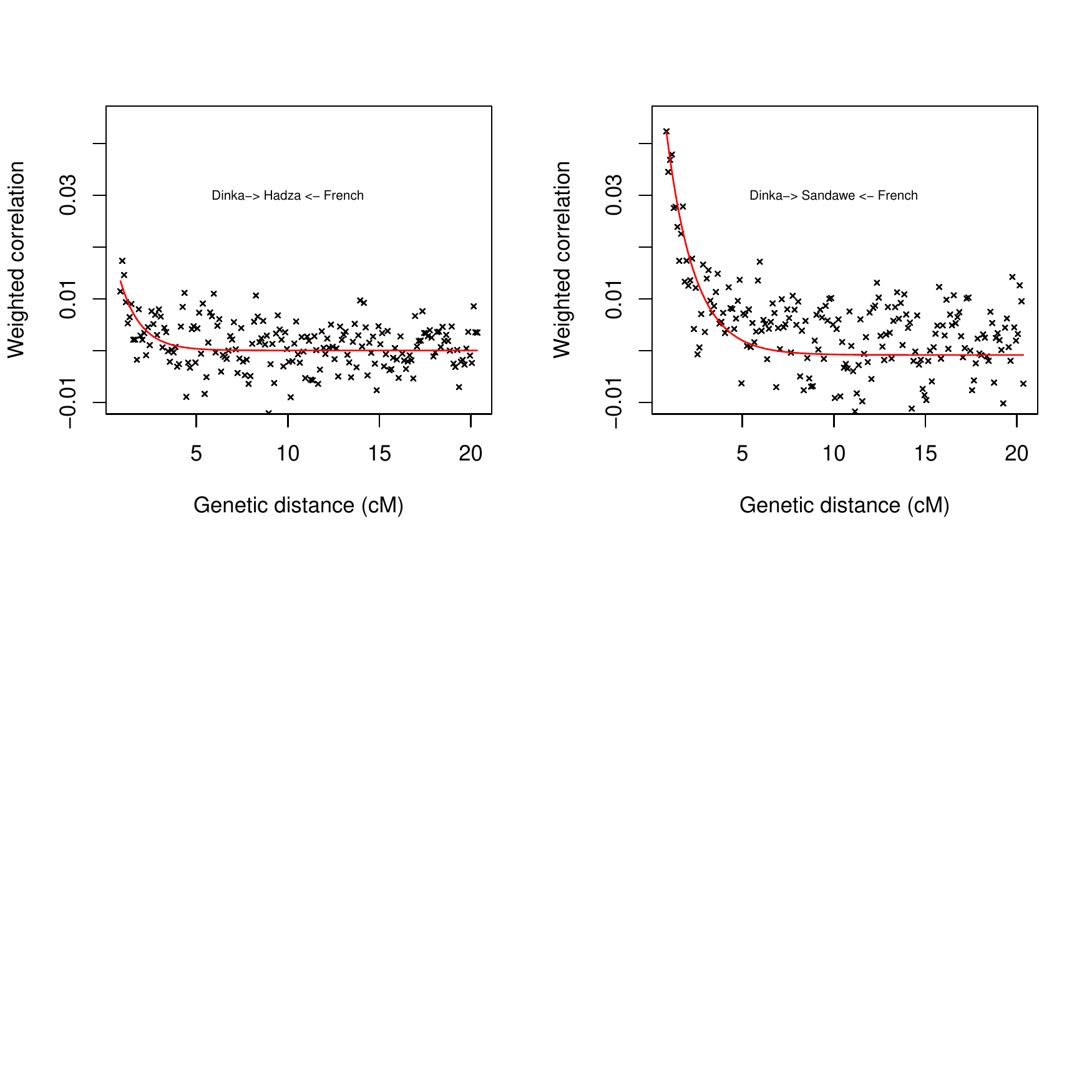}

\caption{\textbf{: Admixture linkage disequilibrium in the Hadza and Sandawe}. We ran ROLLOFF \rolloff on the Hadza and Sandawe, using the Dinka and French as reference populations. Shown are the resulting curves for the Hadza and Sandawe. There is a striking curve of admixture LD in the Sandawe, which is weaker in the Hadza.}\label{rolloff_sandawe}

\end{center}
\end{figure}

\begin{figure}
\begin{center}
\includegraphics[scale = 1]{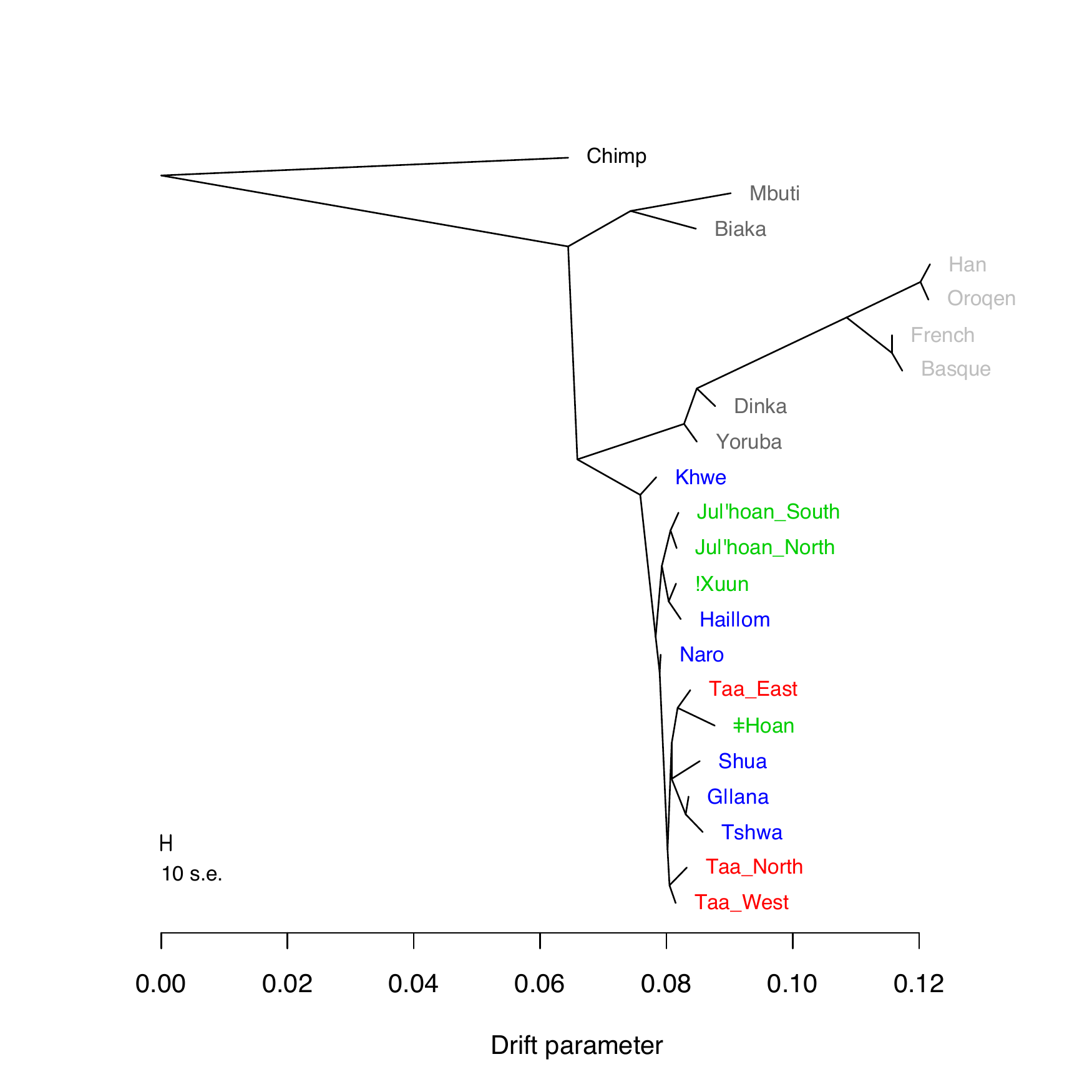}

\caption{\textbf{: \emph{TreeMix} analysis including the Mbuti and Biaka.} We used a modified \emph{TreeMix} approach to build a tree of populations after subtracting out Bantu or Dinka-like ancestry. Shown is the resulting tree; see the Supplementary text for details and discussion.}\label{treemix_pygmy}

\end{center}
\end{figure}




\clearpage

%
%

\section{Supplementary Tables}
\begin{table}[!tbph!]
\begin{center}
\begin{tabular}{ l | c | c | c|}
Population & Language family & Linguistic subgroup & \# of samples \\  \hline
Taa$\_$East & Tuu & Taa-Lower Nossob & 6 \\ 
Taa$\_$North & Tuu & Taa-Lower Nossob &6 \\
Taa$\_$West & Tuu & Taa-Lower Nossob &8\\
!Xuun &Kx'a& Northwest Ju  & 13\\
Ju$|$'hoan\_North & Kx'a & Southeast Ju & 16\\
Ju$|$'hoan\_South & Kx'a & Southeast Ju & 9\\
\textdoublebarpipe Hoan & Kx'a & \textdoublebarpipe Hoan & 7\\
Shua & Khoe-Kwadi & East Kalahari Khoe&  10\\
Tshwa & Khoe-Kwadi &  East Kalahari Khoe &10 \\
Khwe & Khoe-Kwadi & West Kalahari Khoe, Kxoe branch&10\\
Naro & Khoe-Kwadi &  West Kalahari Khoe, Naro branch & 10 \\
G$|$ui & Khoe-Kwadi & West Kalahari Khoe, G$||$ana branch & 5 \\
G$||$ana & Khoe-Kwadi& West Kalahari Khoe, G$||$ana branch  & 5 \\ 
Hai$||$om & Khoe-Kwadi & KhoeKhoe&10\\
Nama & Khoe-Kwadi & KhoeKhoe& 16 \\
Damara  & Khoe-Kwadi & KhoeKhoe&15 \\ 
Kgalagadi & Niger-Congo & Bantu&5\\
Wambo & Niger-Congo & Bantu&5\\
Mbukushu & Niger-Congo & Bantu& 4\\
Tswana & Niger-Congo & Bantu& 5\\
Himba & Niger-Congo & Bantu& 5\\
Hadza & isolate &Hadza& 7\\
\hline
\end{tabular}
\end{center}
\caption{: Summary of samples genotyped in this study.} \label{samp_table}
\end{table}

\clearpage
\begin{table}[!tbp]
\begin{center}
\begin{tabular}{ l | c |}
Sample & Population \\  \hline
BOT6.090 & Ju$|$'hoan\_South \\ 
NAM066 & Ju$|$'hoan\_South \\ 
NAM051 & Ju$|$'hoan\_South  \\ 
BOT6.025 & Taa\_North \\
BOT6.255 & Shua \\
NAM189 & !Xuun\\
NAM195 & !Xuun \\
BOT6.004 & Kgalagadi \\
DR000071 & Hadza  \\
BOT6.058 & Naro  \\
\hline
\end{tabular}
\end{center}
\caption{: Individuals removed from analysis.}  \label{filter_table}
\end{table}

\clearpage
\begin{table}[!tbp]
\begin{center}
\begin{tabular}{ l | c | c|  c|}
Target Population & ``Mixing" populations & Minimum $f_3$ & Z-score \\  \hline
Khwe & Ju$|$'hoan\_North, Yoruba & -0.005 & -38.7\\
Hai$||$om & Ju$|$'hoan\_North, Yoruba & -0.005 &  -33.9 \\
Tshwa & Ju$|$'hoan\_North, Yoruba & -0.005 & -29.9\\
Shua & Ju$|$'hoan\_North, Yoruba & -0.004 &-28.8\\
Tswana & Yoruba, Taa\_West  &-0.004 & -23.8 \\
!Xuun & Ju$|$'hoan\_North, Yoruba & -0.004&  -20.3 \\
G$||$ana & Ju$|$'hoan\_North, Yoruba & -0.005 &  -21.3 \\
Kgalagadi & Ju$|$'hoan\_North, Yoruba & -0.002 & -8.4\\
Naro & Ju$|$'hoan\_North, Taa\_North & -0.0006 & -4.4\\
Mbukushu & Ju$|$'hoan\_North, Yoruba & -0.0008 & -3.9 \\
Taa\_West & Taa\_North, Kgalagadi & -0.0008 &  -3.6\\
Wambo& Ju$|$'hoan\_North, Yoruba & -0.0003 & -1.6 \\

\hline
\end{tabular}
\end{center}
\caption{: \textbf{Three-population tests for treeness.} We performed three-population tests on all possible combinations of populations. Shown are all populations with at least one negative $f_3$ statistic, the names of the putative mixing populations that give rise to the minimum $f_3$ statistic, the value of the statistic, and the Z-score. A Z-score of less than -3 corresponds to a p-value of less than 0.001. The populations labeled as ``mixing" populations are those that give the minimum $f_3$ statistic, and are not necessarily the populations that actually mixed historically.} \label{f3_table}
\end{table}

\begin{table}[!tbp]
\begin{center}
\begin{tabular}{ l | c | c|  c|}
Population & \parbox[c]{2.5cm}{\centering Proportion non-Khoisan ancestry} & \parbox[c]{2 cm}{ \centering Date of mixture (gen.)} \\  \hline
Ju$|$'hoan\_North &  0.06 & 35 \\
Ju$|$'hoan\_South & 0.07 & 35 \\
Naro & 0.09 & 37 \\
Taa\_North & 0.11 & 30 \\
Taa\_West & 0.12 & 10 \\
Taa\_East & 0.17 & 11 \\
\textdoublebarpipe Hoan & 0.23 & 11 \\
!Xuun & 0.25 & 16 \\
G$||$ana & 0.38 & 13 \\
Hai$||$om & 0.43 & 28 \\
Tshwa & 0.48 & 19 \\
Kgalagadi & 0.61 & 23 \\
Shua & 0.62 & 22 \\
Khwe & 0.63 & 25 \\
Tswana & 0.76 & 25 \\
Mbukushu & 0.90 & 14 \\
Damara & 0.90 & 25 \\
Himba & 0.93 &  41 \\
Wambo & 0.93 & NA \\

\hline
\end{tabular}
\end{center}
\caption{: \textbf{Admixture parameters in southern Africa.} We report the admixture proportions and times for each southern African population displayed in Figure 2 in the main text.} \label{admix_table}
\end{table}

\clearpage
\section{Supplementary Methods}

\subsection{Data} 
\subsubsection{Sampling} 
The southern African samples included in this study were collected in various locations in Botswana and Namibia as part of a multidisciplinary project, after ethical clearance by the Review Board of the University of Leipzig and with prior permission of the Ministry of Youth, Sport and Culture of Botswana and the Ministry of Health and Social Services of Namibia. Informed consent was obtained from all donors by carefully explaining the aims of the study and answering any arising questions with the help of translators fluent in English/Afrikaans and the local lingua franca; when necessary, a second translation from the local lingua franca into the native language of a potential donor was provided by individuals within each sampling location. Approximately 2ml of saliva were collected in tubes containing 2ml of stabilizing buffer; DNA was extracted from the saliva with a modified salting-out method \citep{quinque2006evaluation}. 

For the purposes of this study, a minimum of 10 unrelated individuals from each linguistic branch of the three Khoisan language families as given by G\"{u}ldemann \citep{gldemann2008linguist} were selected from the total number of samples collected in the field, as shown in Supplementary Table S1 and Supplementary Figure S\ref{fig_languages}; only the \textdoublebarpipe Hoan branch is represented by fewer than 10 individuals due to the small number of samples available. It should be noted that the ``linguistic subgroup" given in the table does not represent the same level of linguistic relationship for all the populations. The group here called Ju$|$'hoan\_North largely corresponds to what is known as Ju$|$'hoan, often simply referred to as ÒSanÓ in the literature. The Ju$|$'hoan\_South are also known as \textdoublebarpipe Kx'au$||$'en. Since the dialectal boundaries are as yet uncertain and since both groups partly self-identified as Ju$|$'hoan, we here chose geographically defined labels. The HGDP ``San" samples were included with the Ju$|$Õhoan\_North sample, since they clearly stem from this population \schuster and empirically we cannot detect any genetic differentiation between them using the data from this study. For the Khwe, five samples each of the $||$Xokhoe and $||$Anikhoe subgroups were combined, while Damara and Nama individuals were chosen to represent the greatest diversity of traditional subgroups. Since not enough samples were available from all the different Taa dialects, the subgroups of Taa investigated here were chosen based on both linguistic and geographic criteria; they do not correspond to any single linguistic unit. The Taa\_West group includes speakers of the West !Xoon and !Ama dialects, Taa\_North comprises speakers of the East !Xoon dialect, and Taa\_East includes speakers of the Tshaasi and \textdoublebarpipe Huan dialects.  In addition, five samples each from different Bantu-speaking groups from southwestern Zambia (Mbukushu), Namibia (Himba and Wambo), and Botswana (Kgalagadi and Tswana) were included.

The Hadza samples are a subset of those from Henn et al. \henn. Genotypes from other populations were available from other sources, as described below. 

\subsubsection{Genotyping}
Samples were sent to Affymetrix to be genotyped on the Human Origins array. Full details about this array are in Patterson et al. \admixtools, but briefly, SNPs were ascertained by identifying heterozygous SNPs in low-coverage sequencing of single individuals of known ancestry. The SNPs on the array can thus be split into panels of SNPs discovered in different individuals. In all analyses, we consider only autosomal SNPs. Except where otherwise noted, we restrict ourselves to using the 150,425 SNPs discovered in a single Ju$|$'hoan\_North (HGDP ``San") individual. The exception to this are all ROLLOFF analyses (e.g., Figures 2A and 2C in the main text), where we used all 565,259 SNPs on the array. 

The Dinka genotypes were taken from Meyer et al. \denisova. Genotypes from other populations were described in Patterson et al. \admixtools.

\subsubsection{Merging data from \lachance  }
Full genome sequences for five Sandawe, five Hadza, and five Baka/Bakola individuals were obtained via Complete Genomics by \lachance and merged with publicly available Complete Genomics data from a number of other individuals. For our purposes, the most important of these samples are 1) the five Sandawe samples, 2) the five Hadza samples, for comparison to the Hadza we have genotyped, and 3) the HapMap ``YRI" samples, since we have genotyped some of these exact samples on the Affymetrix Human Origins array (and thus can get estimates of genotype error rate). We extracted the genotypes at each SNP on the Human Origins array from the Complete Genomics sequences.

To detect SNPs where genotyping (on either the array or via sequencing) performed poorly, we used a set of eight Yoruba (HapMap YRI) samples that were both sequenced by Complete Genomics and genotyped on the Human Origins array. We removed all SNPs where there were any discordant genotypes between these 8 samples. There were 10,888 such SNPs. We then merged the sequenced Hadza and Sandawe samples into the southern African dataset, again considering only autosomal SNPs. For all analyses involving the Hadza and Sandawe, we included these Hadza and Sandawe individuals. We used the sequenced YRI samples only for quality control.

\paragraph{Quality control.}
Since the Hadza population is quite small, we first used plink \citep{Purcell:2007fk} to test whether the two sets of Hadza samples (those genotyped on the array and those genotyped by sequencing) contained any relatives. We removed one individual that appeared to be the exact same individual in the two samples ($\hat \pi = 0.98$ when using the \texttt{--genome} option in plink)

We then wanted to ensure that there were no systematic differences between samples directly genotyped on the array and those genotyped by sequencing. To look for systematic effects, we used the clustering algorithm ADMIXTURE \citep{Alexander:2009fk}. Two populations, the Hadza and the Yoruba/YRI, include some samples genotyped on the
array and some genotyped via sequencing. For both of these populations, there do not appear to be any substantial differences between samples typed using the two methods (Section \ref{admix_section}, Supplementary Figure S\ref{southaf_admix_wtish})

\subsubsection{Filtering ``outlier" individuals}

As described in the main text, for analyses where we grouped individuals into populations, we removed genetic outliers. To identify individuals that were genetic outliers with respect to their population, we performed PCA on the genotype matrix using the SNPs ascertained in a single Ju$|$'hoan\_North individual (see Section \ref{cluster_section}). We examined each population in turn, and removed individuals that appear as outliers in their population (Supplementary Figure S\ref{outliers}).  A list of all the individuals removed from subsequent analyses is in Supplementary Table S\ref{filter_table}. We furthermore excluded the G$|$ui population, for whom we could not identify a clear genetic cluster of individuals, since three samples clustered with the southeast Kalahari groups, and two with the Nama and G$||$ana.

\subsection{Clustering analyses} \label{cluster_section}
We performed clustering analysis of the genotype matrix using both PCA \smartpca and ADMIXTURE \citep{Alexander:2009fk}. The latter is a fast implementation of the admixture model of STRUCTURE \citep{Pritchard:2000zr} appropriate for genome-wide data.

\subsubsection{PCA}
We first performed PCA \smartpca  using the SNPs ascertained in the Ju$|$'hoan\_North individual and including some non-African populations (Supplementary Figure S\ref{fig_pca_wnonaf}). Nearly all of the Khoisan fall along a cline between the least admixed Khoisan populations and the rest of the African populations. The one major exception is the Nama, who are scattered in the PCA plot, indicating differential relatedness to non-African populations. We examine this further in Section \ref{nama_section}. 

We then considered only the African populations (excluding the Hadza and Sandawe, as we analyze them separately in Sections \ref{hadza_section} and \ref{sandawe_section}). As described in the main text, we performed PCA using SNPs ascertained in either a Ju$|$'hoan\_North (HGDP ``San" individual) (Supplementary Figure S\ref{fig_pca_nooutliers}A), a Yoruba (Supplementary Figure S\ref{fig_pca_nooutliers}B), or a French (Supplementary Figure S\ref{fig_pca_nooutliers}C).   

\paragraph{Substructure within Khoisan populations.}

The PCA (Figure 1B in main text, or Supplementary Figure S\ref{fig_pca_nooutliers}A) indicates that the Taa$\_$West and the Hai$||$om are genetically substructured populations: Half of the Taa$\_$West individuals fall into the southeast Kalahari cluster, while the other half cluster with Nama and G$||$ana. This genetic substructure correlates with a major linguistic boundary: the individuals falling into the southeast Kalahari cluster speak the West !Xoon dialect of Taa, while three of the other individuals speak the !Ama dialect. In the case of the Hai$||$om, four individuals cluster close to the Tshwa and Khwe, while the other five cluster with the !Xuun. This genetic substructure in the Hai$||$om reflects geographic variation: the five individuals that show affnities with the !Xuun come from close to the Angolan border in northern Namibia, where the two groups are settled in close proximity, while the remaining Hai$||$om come from the Etosha area further to the southwest. Conversely, for some groups the known ethnolinguistic subdivisions do not correspond to genetic distinctions. Thus, the Damara sample included four individuals from Sesfontein, and the Nama sample included five Topnaar from the Kuiseb Valley; these groups are linguistically distinct \citep{haacke2008linguistic}, but appear genetically indistinguishable from other Damara and Nama, respectively. Similarly, the two Khwe subgroups cannot be distinguished from each other in the analyses.

\paragraph{PCA projection.}
In Supplementary Figure S\ref{fig_pca_nooutliers}C, there is a shift of some Khoe-speaking populations on the y-axis. For the Nama, we show in Section \ref{nama_section} that this is due to recent European ancestry. For the other populations, we speculate that this may be due to eastern African ancestry. We performed a PCA projection where we first ran the analysis using only the Ju$|$'hoan\_North, Yoruba, and Dinka, and then projected the remaining Khoisan populations on the identified PCs. In this analysis, the Yoruba represent western African populations and the Dinka represent eastern African populations. This analysis was performed on the French-ascertained SNPs, as these are the SNPs where a potential eastern African signal in some of the southern Africans is seen in Supplementary Figure S\ref{fig_pca_nooutliers}C. The results are shown in Supplementary Figure S\ref{fig_pca_project}. The projected samples fall on a line between their Bantu-speaking neighbors and the Ju$|$'hoan\_North. This suggests that the majority of the variation in admixture in these populations is due to variable levels of admixture with their neighbors. However, we cannot rule out some level of admixture with non-Bantu-speaking populations. 

\subsubsection{ADMIXTURE analyses} \label{admix_section}
We ran ADMIXTURE \citep{Alexander:2009fk} on all of the individuals in the African populations (including the French, Basque, Han and Oroqen as reference non-African populations). To prepare the data for analysis, we thinned SNPs in LD using plink \citep{Purcell:2007fk}, as suggested by the authors \citep{Alexander:2009fk}. The precise command was \texttt{--indep-pairwise 50 10 0.1}. Results for different numbers of clusters are shown in Supplementary Figure S\ref{southaf_admix_wtish} (this is for the combined set of southern and eastern African populations). We recapitulate previous results showing that clustering analyses find correlations in allele frequencies between southern African populations and the Mbuti, Biaka, Hadza, and Sandawe \tishkoff (see K = 4). We additionally recapitulate (at K = 8) the PCA result showing detectable structure between northwestern and southeastern Kalahari populations.

\subsubsection{European ancestry in the Nama} \label{nama_section}
The positions of the Nama individuals in Supplementary Figure S\ref{fig_pca_wnonaf} are suggestive of post-colonial European admixture, in accordance with historic documentation of European ancestry in some Nama groups \wallace. To test this, we used four-population tests \india of the form [[Yoruba, X],[Han, French]], where X is any southern African population. A positive $f_4$ statistic indicates gene flow between X and a population related to the French (or alternatively gene flow between populations related to the Yoruba and Han). The most strongly positive $f_4$ statistic in the southern African populations is for the Nama (Supplementary Figure S\ref{fig_mix_eu}A), as expected if they have experienced European admixture. To confirm the direction of this gene flow, we used ROLLOFF \rolloff to test if there is detectable admixture LD in the Nama. If we use the Ju$|$'hoan\_North and the French as the putative mixing populations, there is clear admixture LD (Supplementary Figure S\ref{fig_mix_eu}B), we date this mixture to approximately five generations ($\approx$150 years) ago. The single exponential curve does not perfectly fit the curve in the Nama at shorter genetic distances (Supplementary Figure S\ref{fig_mix_eu}B), indicating that they were likely admixed with some non-Khoisan group at the time of the European admixture. This means that the extremely recent inferred date of admixture in the Nama (five generations) may still be a slight overestimate. 

Interestingly, a few of the Khoe-speaking populations have slightly positive $f_4$ statistics in this comparison, and in the Shua the $f_4$ statistic is significantly greater than zero. We speculate that some of the Khoe-speaking populations have a low level of east African ancestry, and that the relevant east African population was itself admixed with a western Eurasian population. The Shua also show a detectable signal of admixture LD, though we estimate the admixture date as much older (44 generations). This potential signal of potential east African ancestry specifically in Khoe-speaking populations is of particular interest in the light of the hypothesis that the Khoe-Kwadi languages were brought to southern Africa by a pre-Bantu pastoralist immigration from eastern Africa \citep{gldemann2008linguist}. 

Given that the Nama are the only pastoralist Khoisan group included in our dataset, their relationship to the other Khoisan populations is of particular interest. Unfortunately, the recent European admixture they have undergone prevents us from including them in further analyses. However, as shown by the PCA based on Ju$|$Õhoan SNPs (Fig. 1B in main text), the Nama do not stand out among the other Khoisan populations, notwithstanding their distinct life-style. Rather, they cluster closely with Tshwa and G$||$ana foragers, who also speak languages belonging to the Khoe-Kwadi family, on a cline leading to the southeastern Kalahari cluster. To what extent this genetic proximity of the Nama to foraging groups is due to extensive admixture between immigrating pastoralists and resident foragers  \citep{gldemann2008linguist} or rather to a cultural diffusion of pastoralism to indigenous hunter-gatherers \kinahan cannot be addressed at this point. 

\subsection{Three- and four-population tests}
Three- and four-population tests for admixture are described most thoroughly in Reich et al. \india and Patterson et al. \admixtools. We used the implementation of $f_3$ and $f_4$ statistics available as part of the \emph{TreeMix} package \treemix. In all cases standard errors for $f-$statistics were calculated in blocks of 500 SNPs (i.e. -K 500). 

A significantly negative $f_3$ statistic is evidence for admixture in the tested population. We performed all possible three-population tests on the southern African dataset after removing outliers; all populations with negative $f_3$ statistics are shown in Supplementary Table S\ref{f3_table}. Note that genetic drift since admixture reduces the power of this test \india. 

In various places, we use four-population tests. In these tests, of the form $f_4(A,B; C,D)$  (where $A$, $B$, $C$, and $D$ are populations) a significantly positive statistic indicates gene flow between populations related to either $A$ and $C$ or $B$ and $D$, and a significantly negative statistic indicates gene flow between populations related to $A$ and $D$ or $B$ and $C$. 

\subsection{Using the decay of linkage disequilibrium to test for historical admixture} \label{ld_section}
\subsubsection{Motivation}
A common approach to looking for historical admixture in a population is to use clustering analyses like those implemented in STRUCTURE \citep{Pritchard:2000zr} and PCA \smartpca. These are useful approaches to summarizing the major components of variation in genetic data. More formally, these approaches attempt to model the genotypes of each sampled individual as a linear combination of unobserved allele frequency vectors. These vectors (and the best linear combination of them for approximating the genotypes of each individual) are then inferred by some algorithm. PCA and STRUCTURE-like approaches differ only in how the approximation is chosen \citep{Engelhardt:2010fk}. In applications to population history, the inferred allele frequency vectors are often interpreted as ``ancestral" frequencies from some set of populations (in the STRUCTURE-like framework), and the linear combination leading to an individual's genotype as ``admixture" levels from each of these populations. However, the inferred populations need never have existed in reality. Consider two historical scenarios: 1) an individual with 50\% ancestry from a population with an allele frequency of 1 and 50\% ancestry from a population with an allele frequency of 0; and 2) an individual with 100\% ancestry from a population with an allele frequency of 0.5. From the point of view of a clustering algorithm, these two scenarios are identical.

The above hypothetical situation provides some intuition for situations where clustering approaches might mislead. Consider the situation depicted in Supplementary Figure S\ref{admix_sim}A. Here, there are two populations that split apart 3,200 generations in the past. Then, 40 generations in the past, 10\% of one of the populations was replaced by the other (the simulation command is given in Section \ref{sim_section}). We now sample 20 individuals from each population in the present day and run ADMIXTURE \citep{Alexander:2009fk}. With the above intuition, it is not surprising that the algorithm does not pick up the simulated admixture event (Supplementary Figure S\ref{admix_sim}B,D). 

Our goal here is to find a method that does detect admixture in this simple situation, and to estimate the admixture proportions. To do this, we will use the decay of linkage disequilibrium (LD) rather than the allele frequencies alone. Some aspects here are motivated by clustering approaches that use LD information \citep{Falush:2003uq, Lawson:2012fk}, and a related approach is taken by Myers et al. \citep{Myers:2011z}. 

\subsubsection{Methods}

Consider a population $C$, which has ancestry from two populations ($A$ and $B$) with admixture proportions $\alpha$ and $1-\alpha$. Now consider two loci separated by a genetic distance of $x$ cM, and let the allele frequencies at these loci in population $A$ be $f_1^A$ and $f_2^A$, respectively. Define $f_1^B$, $f_2^B$, $f_1^C$, and $f_2^C$ analogously. In a given population (say $A$), define the standard measure of linkage disequilibrium $D_{12}^A = f_{12}^A -  f_1^A f_2^A$, where $f_{12}^A$ is the frequency of the haplotype carrying both alleles 1 and 2 in population $A$. Suppose populations $A$ and $B$ are in linkage equilibrium, so that $D_{12}^A = D_{12}^B = 0$. Now let $C$ result from admixture between populations $A$ and $B$, and for the moment assume infinite population sizes. At time $t$ generations after admixture, then \citep{Chakraborty:1988fk}:

\begin{equation*}
\stepcounter{equation}
\tag{S{\theequation}}
D_{12}^C(t) = \alpha (1-\alpha) e^{-tx} [f_1^A - f_1^B][ f_2^A - f_2^B].
\end{equation*}

\noindent Since $f_1^C = \alpha f_1^A + (1-\alpha) f_1^B$, we can now write $f_1^B = \frac{f_1^C - \alpha f_1^A}{1-\alpha}$. We thus have:

\begin{equation}\label{roll_eq}
\stepcounter{equation}
\tag{S{\theequation}}
D_{12}^C(t) = \frac{\alpha}{1-\alpha} e^{-tx} [f_1^A - f_1^C][ f_2^A - f_2^C]
\end{equation} 

Now assume we have genotyped $m$ SNPs in $n_A$ haplotypes from population $A$ and $n_C$ haplotypes from population $C$. We need to estimate both allele frequencies and linkage disequilibrium in $C$; we do this by splitting the population in half. Let $\hat D_{ij}$ be the estimated amount of linkage disequilibrium between SNPs $i$ and $j$ in population $C$ (using one half of the population), $\hat f_i^C$ be the estimated allele frequency of SNP $i$ in population $C$ (from the other half of the population), $\hat f_i^A$ be the estimated allele frequency of SNP $i$ in population $A$ (in practice a population closely related to $A$ rather than $A$ itself), and $\hat \delta_i$ be $\hat f_i^A - \hat f_i^C$. We now split pairs of SNPs into bins based on the genetic distance between them. We use a bin size of 0.01 cM. In each bin, we calculate the regression coefficient $\hat \beta_x$ from a regression of $\hat D$ on $\hat \delta_i \hat \delta_j$ .  If we let $s$ be the set of all pairs $\{i,j\}$ of SNPs in bin $x$, then

\begin{equation*}
\stepcounter{equation}
\tag{S{\theequation}}
\hat \beta_x = \frac{ \sum \limits_{ \{i,j\} \in s} \hat \delta_{i}\hat \delta_j \hat D_{ij} }{  \sum \limits_{ \{i,j\} \in s}\hat \delta_{i}^2  \hat \delta_j^2 }. 
\end{equation*}

This is a downwardly-biased estimate of $\beta_x$ due to finite sample sizes, since $E[\hat \delta_i^2] \ne \delta_i^2$. To correct for this, note that $\delta_i^2$ is simply an $f_2$ statistic \india, and $\hat \delta_i^2$ is the biased version of the $f_2$ statistic. Now call $\hat f_2$ the biased estimate of the $f_2$ distance between A and C and $\hat f_2^{\star}$  the unbiased estimate of this distance (from Reich et al. \india). We can calculate a corrected version of the regression coefficient, which we call $\hat \beta_x^\star$:

\begin{equation*}
\stepcounter{equation}
\tag{S{\theequation}}
\hat \beta_x^\star = \hat \beta_x \frac{\hat f_2^2}{\hat f_2^{\star2}}.
\end{equation*}

Now, returning to the population genetic parameters, recall that (from Equation \ref{roll_eq}):

\begin{equation*}
\stepcounter{equation}
\tag{S{\theequation}}
\beta_x = \frac{\alpha}{1-\alpha} e^{-tx}. 
\end{equation*}
\noindent We thus fit an exponential curve to the decay of the regression coefficient with genetic distance using the nls() function in R \citep{R-Development-Core-Team:2011uq}.  To remove the effects of LD in the ancestral populations, we ignore distance bins less than 0.5 cM. The amplitude of this curve is an estimate of $\frac{\alpha}{1-\alpha}$, and the decay rate is an estimate of $t$. The interpretation of the amplitude in terms of the admixture proportion relies heavily on the assumption that population $A$ has experienced little genetic drift since the admixture event, and so may not be applicable in all situations (we show below via simulations that this approximation performs well in a situation like that of the Khoisan). 

\subsubsection{Simulations} \label{sim_section} The above theory is valid in the absence of drift and in the presence of phased haplotypes. To test how well this works in more realistic situations, we performed coalescent simulations using macs \citep{Chen:2009fk}. We simulated two populations that diverged 3,200 generations ago, each of which has an effective population size of 10,000. One population then mixes into the other 40 generations ago with some admixture fraction $\alpha$. The parameters were chosen to be reasonable for the Yoruba and Ju$|$'hoan\_North. We simulated $\alpha$ of 0, 0.1, and 0.2. The macs parameters were (for e.g., $\alpha = 0.1$):\\
\\
\texttt{macs 80 100000000 -t 0.0004 -r 0.0004 -I 2 40 40 -em 0.001 2 1 4000 -em 0.001025 2 1 0  -ej 0.08 2 1}
\\

To mimic the effects of uncertain phasing, we randomly combined the simulated chromosomes into diploids, and re-phased them using fastPHASE \citep{Scheet:2006ri}. We then used the above model to estimate the admixture proportions. We simulated five replicates of each $\alpha$, and averaged the resulting curves (Supplementary Figure S\ref{rolloff_sim}). We see that the curves are approximately those predicted by theory, though they slightly overestimate the true mixture proportions. At higher mixture proportions (30\% or 40\%), phasing errors become a major problem and $\alpha$ is severely underestimated (not shown). Two representative simulations of $\alpha = 0.1$ are shown in Supplementary Figures \ref{admix_sim}C,E for comparison to the results from ADMIXTURE. 

\subsubsection{Application to the Khoisan} We then applied this procedure to the five Khoisan populations that do not show significant evidence for admixture from three-population tests (Supplementary Table S\ref{f3_table}). These are the Ju$|$'hoan\_North, Ju$|$'hoan\_South, \textdoublebarpipe Hoan, Taa\_North, and Taa\_East. We phased the merged dataset using fastPHASE, combining all populations (using 20 states in the HMM; i.e. $K=20$). Genetic distances between SNPs were taken from the HapMap \citep{Myers:2005bh} (all genetic maps are highly correlated at the scale we are considering). We used the Yoruba as a reference non-Khoisan population, and use the admixed population itself as the other reference (as described in the theory presented above). All LD decay curves for these populations are shown in Supplementary Figure S\ref{rolloff_noadmix}. All five Khoisan populations show a clear curve; we estimate that the Ju$|$'hoan\_North are the least admixed population, with approximately 6\% non-Khoisan ancestry. 

A potential concern is that demographic events other than admixture (like population bottlenecks) may also lead to substantial LD in some populations. This concern arises because we use the Khoisan population twice in Equation \ref{roll_eq} (population $C$)--both to calculate allele frequencies and to calculate linkage disequilibrium. Though we have used different individuals for these two calculations, there could be unmodeled relationships between the individuals in the two sets. To test the robustness of the curves, we used ROLLOFF \rolloff. In ROLLOFF, the target population is used only to calculate LD, and two other populations are used as representatives of the putative mixing populations; see Moorjani et al. \rolloff for details. While demographic effects in the target population may influence LD, under the null model that the target population is unadmixed, the influence on LD will not be correlated to differences in allele frequencies between two unrelated populations. Results for using the Ju$|$'hoan\_North as the target population and various other pairs of populations as the mixing populations are shown in Supplementary Figure S\ref{rolloff_ju}. There is a clear exponential decay of LD in nearly all cases. For example, the level of LD between two distant SNPs in the Ju$|$'hoan\_North is correlated with the divergence of those SNPs between the Yoruba and the French (Supplementary Figure S\ref{rolloff_ju}); this is not expected if the Ju$|$'hoan\_North are unadmixed. 

\subsubsection{The $f_4$ ratio test in the presence of admixed ancestral populations}
The $f_4$ ratio test was introduced in Reich et al. \india as a method to estimate mixture proportions in an admixed group. In our case, imagine we had samples from Chimpanzee (C), Dinka (D), Yoruba (Y), an unadmixed Khoisan population (S), and an admixed Khoisan population (X). In this setup, the chimpanzee is an outgroup, the Yoruba and population ``S" represent populations related to the admixing populations, and the Dinka are a population that split from the Yoruba before the admixture. Following the derivation in Reich et al. \india, if we let $\alpha$ be the amount of Yoruba-like ancestry in population X:

\begin{equation*}
\stepcounter{equation}
\tag{S{\theequation}}
\frac{f_4(C,D; X, Y)}{ f_4(C, D; S, Y)} = 1- \alpha
\end{equation*}
\noindent However, we do not have samples from S; instead, we \emph{only} have samples from admixed Khoisan populations. Now let $\alpha_1$ be the fraction of Yoruba-like ancestry in population X, and $\alpha_2$ be the fraction of Yoruba-like ancestry in population $S$. If we assume the Yoruba-like mixture into X and S occurred from the same population, then:

\begin{equation*}
\stepcounter{equation}
\tag{S{\theequation}}
\frac{f_4(C,D; X, Y)}{ f_4(C, D; S, Y)} = \frac{1-\alpha_1}{1-\alpha_2}
\end{equation*}
\noindent so

\begin{equation*}
\stepcounter{equation}
\tag{S{\theequation}}
\alpha_1 = 1- (1-\alpha_2)\frac{f_4(C,D; X, Y)}{ f_4(C, D; S, Y)}.
\end{equation*}
\noindent Of course, using this approach requires an independent method for calculating $\alpha_2$. We use the Ju$|$'hoan\_North as population S, and estimate $\alpha_2$ from the linkage disequilibrium patterns as described in the previous section. 

\subsection{Estimating mixture dates with ROLLOFF}
We used ROLLOFF  \rolloff to estimate admixture dates for all southern African populations. To do this we set the Ju$|$'hoan\_North and Yoruba as the two mixing populations (note that the date estimates in ROLLOFF are robust to improper specification of the mixing populations \rolloff), and ran ROLLOFF on each population separately (Supplementary Figure S\ref{rolloff}). We generated standard errors on the date estimates by performing a jackknife where we drop each chromosome in turn, as in Moorjani et al. \rolloff. In this analysis, we used all the SNPs on the genotyping chip, and genetic distances from the HapMap \citep{Myers:2005bh} (all genetic maps are highly correlated at the scale we are considering). For the Ju$|$'hoan\_North, we used half the sample to estimate allele frequencies and half to estimate LD, as in Section \ref{ld_section}. 

ROLLOFF estimates a single date of admixture, while in reality there may be multiple waves (or continuous) admixture in the history of a population. In these scenarios, the rate at which admixture LD decays is no longer an exponential curve, but instead a \emph{mixture} of multiple exponential curves with different decay rates. We thus examined the residual fit from fitting a single exponential curve to the LD decay in each population (Supplementary Figure S\ref{rolloff_resid}). For a few populations, a single exponential curve does not completely describe the data, especially at shorter genetic distances. This implies that for some populations, most notably the !Xuun, G$||$ana, Taa\_East and Taa\_West, there was likely some admixture before the date estimated by ROLLOFF.

\subsection{Estimating population split times}

\subsubsection{Motivation}
Consider two populations, $X$ and $Y$. These populations split at time $t$ generations in the past, and our goal is to estimate $t$ from genetic data (in our case, SNPs). There are two main approaches that have been applied to this problem in the past. The first approach is based on the observation that it is often impossible to write down the probability of seeing genetic data under a given demographic model, but it is quite easy to \emph{simulate} data under essentially any demographic model. It is thus possible to identify demographic parameters which generate simulated data approximately similar to those observed. This is what is now called approximate Bayesian computation (see Pritchard et al. \citep{Pritchard:2000zr}  and Beaumont et al. \citep{Beaumont:2002vn} for a more formal description), and this approach has been applied to estimating split times between populations in a number of applications (e.g. \citep{Wollstein:2010fk, Patin:2009uq, Lohmueller:2009fk}). 

The other approach to this problem does not rely on simulations, but uses an explicit expression for the joint allele frequency spectrum in two populations under a given demographic history \citep{Gutenkunst:2009zr}. The joint allele frequency spectrum is influenced by a number of demographic parameters, including the effective population sizes of the populations, the time at which they split, and other considerations; Gutenkunst et al. \citep{Gutenkunst:2009zr} estimate all of these parameters. 

In both approaches, the demographic history of the populations modeled is assumed to be simple--a constant population size, exponential growth, or bottleneck models (or some combination thereof) are popular due to mathematical convenience. However, true population history is almost certainly more complex than can be modeled. For estimates of population split times, however, the population demography is a nuisance parameter, and we do not wish to estimate it. We will attempt to estimate split times with an approach that, in principle, is valid in situations of arbitrary demographic complexity (with some caveats to come later). The approach we will take is most similar in spirit to Gutenkunst et al. \citep{Gutenkunst:2009zr}, but tailored specifically to our data. The main idea is roughly as follows: after the split of two populations, a given chromosome from one of the populations accumulates mutations at a clock-like rate. We wish to count those mutations, and convert this count to absolute time.

\subsubsection{Methods} \label{split_section}
The demographic setting for the model is presented in Supplementary Figure S\ref{fig_split}A. We have an outgroup population $O$, and two populations whose split time $t$ we wish to estimate, $X$ and $Y$. The population ancestral to $X$ and $Y$ is called $A$. An important modeling assumption is that after populations split, there is no migration between then. Now imagine we have identified a number of sites that are heterozygous in a single individual from $Y$ (in applications later on, this will be the Ju$|$'hoan\_North; recall that this is the exact ascertainment scheme used on the Human Origins array). Looking backwards in time, these are simply all the sites where a mutation has occurred on either of the two chromosomes before they coalesce, and can be split into two groups--the mutations that arose on the lineage to $Y$ (these are the red stars in Supplementary Figure S\ref{fig_split}A) and those that did not (and thus were polymorphic in $A$; these are the black stars in Supplementary Figure S\ref{fig_split}A, and the allelic spectrum in $A$ is shown in Supplementary Figure S\ref{fig_split}B)). Now consider the allele frequencies in $X$. The new mutations that arose on the lineage to $Y$ are of course not polymorphic in $X$, which leads to a peak of alleles with frequency zero in both the ancestral population and in $X$  (Supplementary Figure S\ref{fig_split}B,C). More formally, let $f(x)$ be the allelic spectrum in $A$ conditional on ascertainment in a single individual from $Y$. This spectrum can be split into two parts:

\begin{equation*}
\stepcounter{equation}
\tag{S{\theequation}}
f(x) = \begin{cases} \lambda, & \mbox{if } x = 0 \\ (1-\lambda)g(x), & \mbox{otherwise} \end{cases}
\end{equation*}
\noindent where $\lambda$ is the fraction of SNPs that were non-polymorphic in $A$ (i.e., that arose on the lineage to $Y$), and $g(x)$ is the polymorphic frequency spectrum. The key parameter for our purposes is $\lambda$. If population sizes are constant, $g(x)$ is linear, but in more complex situations can take other forms \citep{Keinan:2007kx}. We assume $g(x)$ is a quadratic of the form $ax^2 + bx + c$. This form is motivated by the fact that the observed allelic spectra are not linear (and thus inconsistent with a constant population size), so we took the next most complicated model, which seems approximately appropriate in practice (see e.g. Supplementary Figure S\ref{fig_ntaa}A).

Now we need to write down the (conditional) allelic spectrum in $X$. To do this, we use the diffusion approximation to genetic drift. Let $\tau$ be the drift length (on the diffusion timescale)  between $A$ and $X$ (we show later on how this can be estimated). Now we can write down the allelic spectrum in $X$, $h(x)$:

\begin{equation*}
\stepcounter{equation}
\tag{S{\theequation}}
h(x) = \int \limits_0^1 f(y)  \kappa^{\star}(x; y, \tau) dy
\end{equation*}
\noindent where $\kappa^{\star}(x; y, \tau)$ is the probability that an allele at frequency $y$ in $A$ is now at frequency $x$ in $X$, given $\tau$. This is closely related to the Kimura transition probability \citep{kimura1955solution}, which we call  $\kappa(x; y, \tau)$. However, the Kimura transition probability is the \emph{polymorphic} transition probability, while we want the probabilities of fixation as well:

\begin{equation*}
\stepcounter{equation}
\tag{S{\theequation}}
\kappa^{\star}(x; y, \tau) = \begin{cases} 1- \int \limits_0^1 \kappa(z; y, \tau) dz) - (y - \int \limits_0^1 z \kappa(z; y, \tau) dz) , & x = 0 \\ \kappa(x;y, \tau), & \mbox{0 $< x <$ 1} \\ y - \int \limits_0^1 z \kappa(z; y, \tau) dz & x = 1 \end{cases}
\end{equation*}

Now we can write down a likelihood for observed data. Let $n$ be the number of SNPs,  let $m_i$ be the number of sampled alleles in $X$ at SNP $i$, and let $c_i^D$ be the number of counts of the derived allele at SNP $i$. Let $c^D$ (with no subscript) be the vector of counts of derived alleles.The likelihood for the data is then:
\begin{equation*}
\stepcounter{equation}
\tag{S{\theequation}}
l(c^D | \lambda, g(x)) = \prod \limits_{i = 1}^n \int \limits_0^1 Bin(c^D_i; m, p) h(p) dp
\end{equation*}
\noindent where $Bin(c^D_i; m, p)$ is the binomial sampling probability. This likelihood can be calculated using numerical integration. We now have three parameters to estimate: two parameters of the polymorphic spectrum in the ancestral population ($a$, and $b$; $c$ is just a scaling factor, which we fix as 1; we normalize the spectrum so that it is a true probability distribution), and $\lambda$. We start with a linear ancestral spectrum and optimize each parameter in turn until convergence. To calculate standard errors of the estimates of these parameters, we perform a block jackknife \india in blocks of 500 SNPs.

\subsubsection{Estimation of $\tau$} \label{tau}
An important parameter in this model is $\tau$, the amount of genetic drift (in diffusion units) that has occurred on the branch from $A$ to $X$. All the complexity of the changes in population size on this branch are absorbed into this parameter. Formally:
\begin{equation*}
\stepcounter{equation}
\tag{S{\theequation}}
\tau = \int \limits_0^t \frac{1}{2N(s)}ds
\end{equation*}
\noindent where $N(s)$ is the effective population size at time $s$ in $X$. To estimate this, we rely on SNPs ascertained by virtue of being heterozygous in a single individual in an outgroup population (in applications later on this will be the Yoruba). At a given such SNP, let the derived allele frequency be $a$ in $A$, $x$ in $X$, and $y$ in $Y$. Now consider the following quantities:
\begin{equation*}
\stepcounter{equation}
\tag{S{\theequation}}
N = x(1-x)
\end{equation*}
\noindent and 
\begin{equation*}
\stepcounter{equation}
\tag{S{\theequation}}
D = x(1-y).
\end{equation*}
\noindent Now consider the expectations of these quantities. For $N$, this is simply the expected heterozygosity; for a coalescent derivation see Wakeley \citep{wakeley2009coalescent}:
\begin{align*}
\stepcounter{equation}
\tag{S{\theequation}}
E[N] &= E[ x(1-x)] \\
& = a (1-a ) e^{-\tau}
\end{align*}
\noindent and for $D$, since $x$ and $y$ are conditionally independent given $a$:
\begin{equation*}
\stepcounter{equation}
\tag{S{\theequation}}
E[D] = a(1-a).
\end{equation*}

\noindent We now need unbiased estimators of $N$ and $D$. Let there be $n$ SNPs in the panel used for calculating $\tau$, let $\hat x_i$ be the estimated frequency of the derived allele at SNP $i$ in population $X$, and let $\hat y_i$ be the estimated frequency of the derived allele at SNP $i$ in population $Y$.  An estimator of $N$ (which we call $\hat N$) is:
\begin{equation*}
\stepcounter{equation}
\tag{S{\theequation}}
\hat N = B_x + \frac{1}{n}\sum_{i = 1}^n \hat x_i (1- \hat x_i)
\end{equation*}
\noindent where $B_x$ is a correction to make this an unbiased estimator (the calculation for $B_x$ is Equation 4 from the Supplementary Material in Pickrell and Pritchard \treemix). The estimator of $D$ is the trivial one:
\begin{equation*}
\stepcounter{equation}
\tag{S{\theequation}}
\hat D = \frac{1}{n}\sum_{i = 1}^n \hat x_i (1- \hat y_i)
\end{equation*}
We thus have an estimate of $\tau$:
\begin{equation*}
\stepcounter{equation}
\tag{S{\theequation}}
\hat \tau = - \log \bigg(\frac{\hat N}{\hat D}\bigg).
\end{equation*}

\subsubsection{Calibration}
Once we've estimated $\lambda$, we would like to convert this to $t$. $\lambda$ is the proportion of all SNPs ascertained using two chromosomes in population $Y$ that arose on the lineage specific to $Y$ (Figure \ref{fig_split}A), and can thus be written in terms of the \emph{total} number of all mutations specific to these chromosomes (both the red and black mutations in Figure \ref{fig_split}A) and the number of these that arose since $t$ (the red mutations in Figure \ref{fig_split}A).  The former is simply the heterozygosity in population $Y$ (call this $h$), and the latter is $2t\mu$, where $\mu$ is the mutation rate. This assumes that the two chromosomes do not coalesce before $t$, which is a fair assumption in our case where the estimated drift on the Ju$|$'hoan\_North lineage is small. We can thus write:

\begin{equation*}
\stepcounter{equation}
\tag{S{\theequation}}
\lambda = \frac{2 \mu t}{h} 
\end{equation*}
\noindent and so:
\begin{equation*}
\stepcounter{equation}
\tag{S{\theequation}}
t = \lambda \frac{h}{2\mu}.
\end{equation*}
\noindent In practice the ratio of the heterozygosity to the mutation rate must be taken from outside estimates;  see Section \ref{application} for the specific numbers used in our applications in the Khoisan. 

\subsubsection{Simulations}
In this section, we validate the above method using simulations and test its robustness to violations of the model. In particular, in our case there is gene flow from a non-Khoisan group into the Khoisan, so the behavior of this method in such situations is quite important. First, using ms \citep{Hudson:2002ys}, we simulated samples from populations with split times at different depths. All simulations used a demography like that in Figure \ref{fig_split}, with samples from an outgroup that split off 3,200 generations in the past, and two populations whose split time we wish to estimate. The exact ms command used, for a split time of 400 generations in the past, was:
\\
\\
\texttt{ms 60 3000 -t 40 -r 40 50000 -I 3 20 20 20 -ej 0.01 3 2 -ej 0.08 2 1}
\\
\\
For each simulation, we then generated two sets of SNPs: one ascertained by virtue of being heterozygous in a single sample from population $O$ (the outgroup), and one ascertained by virtue of being heterozygous in a single sample from population $Y$. The procedure for estimating the split time is then as follows:
\begin{enumerate}
\item Estimate $\tau$ (the drift from $A$ to $X$) using the SNPs ascertained in $O$ and the method in Section \ref{tau}
\item With $\tau$ fixed, estimate $\lambda$ using the SNPs ascertained in $Y$ and the the method from Section \ref{split_section}
\item Convert the estimated $\lambda$ to generations using the calibration (the mutation rate has been set for the simulation and thus is known, and the heterozygosity is estimated in each simulation)
\end{enumerate}

We performed five simulations each at population split times of 400, 1,200, and 2,000 generations. In all cases, the population split time is well-estimated (Supplementary Figure S\ref{fig_sim}).  We then performed simulations where populations $X$ and $Y$ have experienced some admixture from the ``outgroup". In all cases, we simulated 5\% admixture from $O$ into $Y$, and variable levels of admixture from $O$ into $X$. All admixture occurred 40 generations before present. These numbers were chosen to be appropriate for the Khoisan application. The precise ms command (for 5\% admixture in $X$) is:
\\
\\
\texttt{ms 60 3000 -t 40 -r 40 50000 -I 3 20 20 20 -em 0.002 2 1 2000 -em 0.00205 2 1 0 -em 0.002 3 1 2000 -em 0.00205 3 1 0 -ej 0.01 3 2  -ej 0.08 2 1}
\\
\\
We then performed the exact same estimation procedure to get the split times (Supplementary Figure S\ref{fig_sim_wmig}). In all cases, the admixture leads to overestimation of the split time. This is true even when there is no admixture into $X$ (but only into $Y$). 
\subsubsection{Application to the Khoisan} \label{application}
We then applied this method to date the split of the northwestern and southeastern Kalahari populations (the time of the first split in the southern Africans in Figure 3 in the main text). Some caveats of interpretation here are warranted. First, all the Khoisan populations have some level of admixture with non-Khoisan populations. There is thus no single ``split time" in their history, and any method (like the one used here) that estimates a single such time will actually be estimating a composite of several signals. Second, we have made the modeling assumption that history involves populations splitting in two with no gene flow after the split. More complex demographies are quite plausible, but render the interpretation of a split time nearly meaningless (if populations continue to exchange migrants after ``splitting", they arguably have not split at all). We thus consider strong interpretations of split times estimated from genetic data to be impossible, but we nonetheless find the estimates to be useful in constraining the set of historical hypotheses that are consistent with the data.

For all applications to the Khoisan, the population $Y$ is the Ju$|$'hoan\_North, and $O$ (the outgroup) is the Yoruba. All split times are thus split times between the Ju$|$'hoan\_North and another population. We estimated $\tau$ for each population using the set of SNPs ascertained in the Yoruba, and then estimated $\lambda$ using the set of SNPs ascertained in the Ju$|$'hoan\_North. To convert from $\lambda$ to $t$, we need an estimate of the ratio of the heterozygosity in the Ju$|$'hoan\_North to the mutation rate. We took the estimate of this ratio for the Yoruba from Sun et al. \citep{Sun:2012uq} and then used the fact that the heterozygosity in the Yoruba is 95\% of that in the Ju$|$'hoan \denisova. Specifically, we averaged this ratio across six Yoruba individuals (from Supplementary Table 8 in Sun et al. \citep{Sun:2012uq}) and multiplied by 1.04 (to account for the estimated factor by which the heterozygosity in the Ju$|$'hoan\_North is greater than that in the Yoruba) to get an estimate of $\frac{h}{\mu}$. To get from generations to years, we use a generation time of 30 years \citep{fenner2005cross}. 

The resulting split times are shown in Supplementary Figure S\ref{fig_khoisan}. We plot these as a function of non-Khoisan ancestry, as the latter tends to inflate estimates of the split time (Supplementary Figure S\ref{fig_sim_wmig}). As expected, regardless of the level of admixture, the northwestern Kalahari groups have more recent split times (from the Ju$|$'hoan\_North, who are a northwestern Kalahari group) than the southeastern Kalahari groups. The split time of interest is that with the southeastern Kalahari groups (the red points in Supplementary Figure S\ref{fig_khoisan}). The population with the least non-Khoisan ancestry is the Taa\_North; we show the empirical and fitted allele frequency spectra for this population in Supplementary Figure S\ref{fig_ntaa}A, and the simulated allele frequency spectrum from a simulation with an older date of 2,000 generations in Supplementary Figure S\ref{fig_ntaa}B. In the Taa\_North we get a point estimate of 823 $\pm$ 99 generations  ($\approx 25,000 \pm 3,000$ years). However, the simulations in Supplementary Figure S\ref{fig_sim_wmig} indicate that this is likely an overestimate, and perhaps a considerable overestimate. We thus conclude that the split between the northwest and southeast Kalahari groups occurred within the last 30,000 years, and perhaps much more recently than that.

\subsection{\emph{TreeMix} analyses}

\subsubsection{Analysis of the Hadza} \label{hadza_section}
We sought to understand the relationships of the Hadza to the southern African populations. To do this, we selected populations with little admixture to represent the southern African groups (the Taa\_East, Taa\_North, Ju$|$'hoan\_South, and Ju$|$'hoan\_North; see the next section for an analysis of all the Khoisan populations excluding the Damara), African non-Khoisan groups, and non-African groups. We included the chimpanzee sequence as an outgroup. We then built a tree of these populations using \emph{TreeMix} \treemix, which fits a tree to the observed variance-covariance matrix of allele frequencies (Supplementary Figure S\ref{treemix_hadza}A). The Hadza do not group with the southern African populations in this analysis; however, they are poorly modeled by a tree, as seen in the residual fit from the tree (this is the observed covariance matrix subtracted by the covariance matrix corresponding to the tree model; Supplementary Figure S\ref{treemix_hadza}B). 

We then allowed \emph{TreeMix} to build the best graph, allowing for a single admixture event (Supplementary Figure S\ref{treemix_hadza}C). The algorithm infers that the Hadza are admixed between a population related to the southern African Khoisan groups and a population that is most closely related to the Dinka, a northeastern African population. The fraction of Khoisan ancestry in the Hadza is estimated at 23 $\pm$ 2\% (from a block jackknife in blocks of 500 SNPs). The residual fit from this graph is shown in Supplementary Figure S\ref{treemix_hadza}D. The residual covariance of the Hadza with all populations except the Yoruba are less than three standard errors away from the fitted model; for the fit of the covariance between the Yoruba and the Hadza, the fit is 3.5 standard errors away. Indeed, the Yoruba are particularly poorly fitted in this graph, and the worst fit in this graph is for the fit between the Yoruba and the Chimpanzee (Supplementary Figure S\ref{treemix_hadza}D). This poor fit for the Yoruba may indicate archaic admixture (indeed, if we allow \emph{TreeMix} to estimate a second migration edge, it estimates admixture from an archaic population into the Yoruba [not shown]). However, other explanations are possible, and we leave this for future study. 

We compared the \emph{TreeMix} estimate of this Hadza admixture fraction to that obtained by $f_4$ ancestry estimation. We thus calculated $\frac{f_4(Chimp, Yoruba; Hadza, Dinka)}{f_4(Chimp, Yoruba;Ju|'hoan\_North, Dinka)}$, which is an approximation of the fraction of Ju$|$'hoan-like ancestry in the Hadza (though necessarily a slight overestimate due to the non-Khoisan ancestry in the Ju$|$'hoan\_North). This estimate is  27 $\pm$ 1.7\%, which is consistent with the \emph{TreeMix} estimate. To ensure that this estimate is reasonable, we replaced the Hadza by the Bantu-speakers from Kenya from the HGDP (who are an eastern African population not expected to have any Khoisan ancestry) and performed the same analysis. We get an estimate of 3 $\pm$ 1.2\% Khoisan ancestry, confirming the reliability of the estimate.

As discussed in the main text, the major caveat to the interpretation of the Hadza result is that a plausible alternative interpretation for the failure of the tree [Chimp, Ju$|$'hoan\_North, [Dinka, Hadza]] is more archaic gene flow into the ancestors of the Dinka than into the ancestors of the Hadza. There is no signal of Neandertal or Denisova ancestry in the Dinka \denisova, so the source of archaic gene flow would have to be an undiscovered population. We thus prefer the interpretation that the Hadza share ancestry with the Khoisan, though we acknowledge the possibility that future work will challenge this interpretation. 

\subsubsection{Analysis of the Sandawe} \label{sandawe_section}

To begin our analysis of the Sandawe, we performed the same analyses as done with the Hadza. We began by building the maximum likelihood tree of populations including the Sandawe using \emph{TreeMix} (Supplementary Figure S\ref{treemix_sandawe}A). Like the Hadza, the Sandawe are poorly fitted by a tree (Supplementary Figure S\ref{treemix_sandawe}B), so we allowed a single migration edge. The inferred migration event is from a population related to the Khoisan, like we previously saw in the Hadza (Supplementary Figure S\ref{treemix_sandawe}C). The \emph{TreeMix} estimate is that the Sandawe trace about 18\% of their ancestry to a population related to the Khoisan.

\subsubsection{West Eurasian ancestry in the Sandawe and Hadza.} \label{west_sandawe_hadza_section}
We noted that the fit of the Sandawe in the \emph{TreeMix} graph is imperfect (Supplementary Figure S\ref{treemix_sandawe}D). In particular, the relationship between the Sandawe and the European populations in these data is a poor fit. On inspection, the Hadza also show a similar signal, but to a lesser extent (Supplementary Figure S\ref{treemix_hadza}D). We thus examined the Sandawe and Hadza for evidence of west Eurasian ancestry. We used $f_4$ statistics of the form [Chimp, X,[French, Han]], where X is either the Sandawe or the Hadza. In both cases, this tree fails. For the Hadza, this tree fails with a Z-score of -4.2 ($p = 1.3\times 10^{-5}$), and for the Sandawe, this tree fails with a Z-score of -7.2  ($p = 3\times 10^{-13}$). Both of these are consistent with west Eurasian (either European or, more likely, Arabian), gene flow into these populations. To further examine this, we turned to ROLLOFF. We used Dinka and French as representatives of the mixing populations (since date estimates are robust to improperly specified reference populations). The results are shown in Supplementary Figure S\ref{rolloff_sandawe}. Both populations show a detectable curve, though the signal is much stronger in the Sandawe than in the Hadza. The implied dates are 89 generations ($\approx$2500 years) ago for the Hadza and 66 generations ($\approx$2000 years) ago for the Sandawe. These are qualitatively similar signals to those seen by Pagani et al. \citep{Pagani:2012uq} in Ethiopian populations. There are two possible historical scenarios that could lead to these signals: either the Hadza and Sandawe both directly admixed with a western Eurasian population about 2,000 years ago, or they admixed with an east African population that was itself admixed with a western Eurasian population. The latter possibility would be consistent with known east African admixture into the Sandawe \tishkoff. 

\subsubsection{Modification of \emph{TreeMix} to include known admixture}
Since all of the southern African Khoisan populations are admixed with non-Khoisan populations, any attempt to build a tree relating these populations is complicated by admixture. We wanted to examine the historical relationships of these populations before the admixture. To do this, we used the composite likelihood approach of Pickrell and Pritchard \treemix, as implemented in the software \emph{TreeMix}. Briefly, the approach is to build a graph of populations (which allows for both population splits and mixtures) that best fits the sample covariance matrix of allele frequencies \treemix. In all analyses, we calculate the standard errors on the entries in the covariance matrix in blocks of 500 SNPs. 

In the original \emph{TreeMix} algorithm, one first builds the best-fitting tree of populations. However, this approach is not ideal if there are many admixed populations (as in our application here, where all of the Khoisan populations are admixed). To get around this, we allow for \emph{known} admixture events to be incorporated into this tree-building step. Imagine that there are several populations that we think \emph{a priori} might be unadmixed (in our applications, these are the Chimpanzee, Yoruba, Dinka, Europeans, and East Asians). We first build the best tree of these unadmixed populations using the standard \emph{TreeMix} algorithm. Now assume we have an independent estimate of the admixture level of each Khoisan population, and imagine we know the source population for the mixture.

To add a Khoisan population to the tree, for each existing branch in the tree, we put in a branch leading to the new population. We then force the known admixture event into the graph with a fixed weight, update the branch lengths, and store the likelihood of the graph. After testing all possible branches, we keep the maximum likelihood graph. We then try all possible nearest-neighbor interchanges to the topology of the graph (as in the original \emph{TreeMix} algorithm), keeping the change only if it increases the likelihood. We do this for all populations. Finally, after adding all the populations with fixed admixture weights, we optimize the admixture weights, and attempt changes to the graph structure where the source populations for the admixture events are changed.

To initialize the migration weights for each Khoisan population, we used the corrected $f_4$ ratio estimates from Figure 2B in the main text. To initialize the source population for the mixture events, we chose the Yoruba for all populations except the Hadza and Sandawe, which we initialized as mixing with the Dinka. We additionally initialized the Hadza and Sandawe as having 5\% and 10\% ancestry, respectively, related to the French, for the reasons described in Section \ref{west_sandawe_hadza_section} (these proportions were chosen based on rough examination of the ADMIXTURE plot (Supplementary Figure S\ref{southaf_admix_wtish}), but are only used for intialization; the algorithm then updates these proportions to fit the data. The final estimated proportions are 13\% and 17\%, respectively. 

To obtain a measure of confidence in the resulting tree, if there are $K$ blocks of 500 SNPs, we performed a bootstrap analysis where we randomly sample $K$ blocks from the genome (with replacement) and re-estimate the tree. We ran this bootstrap analysis 100 times, then counted the fraction of replicates supporting each split in the tree.

\subsubsection{Analysis of the Mbuti and Biaka}
It has been proposed that the Mbuti and Biaka hunter-gatherers from central Africa were once part of an Africa-wide hunter-gatherer population \tishkoff. We thus tested whether these populations share the same signal of relatedness to the Khoisan as we see in the eastern Africans. We started by looking at $f_4$ statistics of the form [Chimp, Ju$|$'hoan\_North, [X, Dinka]], where X is either the Mbuti or the Biaka. Recall that in the main text we show that, if X is the Hadza or Sandawe, this tree fails in a way that implies admixture between southern and eastern Africa. Here, these trees also fail (for Biaka,  Z = 6.4, p = $7.7\times 10^{-11}$; for Mbuti, Z = 3.1, p =  0.001). However, these failures are in the \emph{opposite} direction to those seen for the eastern Africans. Instead, these $f_4$ statistic imply either archaic ancestry in the Mbuti and Biaka, or gene flow between the Khoisan and the Dinka. We examined whether this signal is robust to the choice of ascertainment panel, instead using the SNPs ascertained in a single Yoruba individual. In this panel, we see the same signals (for Biaka, Z = 9.6, p $< 1\times 10^{-12}$ ; for Mbuti, Z = 6.8, p = 5 $\times 10^{-12}$). The signal of relatedness of the central Africans to the Khoisan is thus qualitatively and quantitatively different than that seen for the eastern Africans. 

We used the approach in the previous section to build a tree relating the Mbuti and Biaka to the Khoisan, like that done in Figure 3 in the main text. As before, we initialized the Khoisan population as having a fraction of their ancestry related to the Yoruba. We initialized the Mbuti and Biaka as having 37\% and 53\%, respectively, of their ancestry related to the Yoruba (these fractions are only used for initialization, but are then updated in the algorithm). The resulting tree is shown in Supplementary Figure S\ref{treemix_pygmy}. As expected based on the $f_4$ statistics, the Mbuti and Biaka do not fall on the same branch as the Khoisan, but instead appear to descend from a population that is an outgroup to everyone else.

\pagebreak

\clearpage
\bibliography{bib}
\end{document}